%% file: StabilizingSqueezedMWLight.tex
\documentclass[12pt]{iopart}
\usepackage{amssymb}
\usepackage{graphicx}
\usepackage{braket}
\usepackage[export]{adjustbox}
\usepackage{cite}

\newcommand{\sbra}[1]{\langle\langle #1|}
\newcommand{\sket}[1]{|#1\rangle\rangle}
\newcommand{\sbraket}[2]{\langle\langle#1|#2\rangle\rangle}

\newcommand{\ann}{\hat{a}}
\newcommand{\adag}[1][]{\hat{a}^{\dagger #1}}

\newcommand{\bnn}{\hat{b}}
\newcommand{\bdag}[1][]{\hat{b}^{\dagger #1}}

{\tiny }\begin{document}

\title{Quantum microwaves: stabilizing squeezed light by phase locking}

\author{
Lukas Danner$^{1,2 *}$,
Florian H\"ohe$^{2 *}$,
Ciprian Padurariu$^2$,
Joachim Ankerhold$^2$ and 
Bj\"orn Kubala$^{1,2}$
}

\address{$^1$Institute for Quantum Technologies, German Aerospace Center (DLR), 89081 Ulm, Germany}
\address{$^2$Institute for Complex Quantum Systems and IQST, University of Ulm, 89069 Ulm, Germany}

\address{$^*$These two authors contributed equally to this work.}
\ead{lukas.danner@uni-ulm.de, florian.hoehe@uni-ulm.de}
\vspace{10pt}

\begin{abstract}
\input{abstract}
\end{abstract}

\input{text_main}

\newpage
\appendix
\input{text_supp}

\section*{References}
\bibliographystyle{unsrt}
\bibliography{references}

\end{document}

%% file: abstract.tex
Bright sources of quantum microwave light are an important building block for various quantum technological applications. Josephson junctions coupled to microwave cavities are a particularly versatile and simple source for microwaves with quantum characteristics, such as different types of squeezing. Due to the inherent nonlinearity of the system, a pure dc-voltage bias can lead to the emission of correlated pairs of photons into a stripline resonator.
However, a drawback of this method is that it suffers from bias voltage noise, which disturbs the phase of the junction and consequently destroys the coherence of the photons, severely limiting its applications. Here we describe how adding a small ac reference signal either to the dc-bias or directly into the cavity can stabilize the system and counteract the sensitivity to noise.
We first consider the injection locking of a single-mode device, before turning to the more technologically relevant locking of two-mode squeezed states, where phase locking preserves the entanglement between photons.
Finally, we describe locking by directly injecting a microwave into the cavity, which breaks the symmetry of the squeezing ellipse. In all cases, locking can mitigate the effects of voltage noise, and enable the use of squeezed states in quantum technological applications.

%% file: text_main.tex
\section{Introduction}
Squeezed states of light are an important resource for various quantum technological applications. Most prominently in metrology they can be used for quantum enhanced precision measurements \cite{xiao1987precision,grangier1987squeezed}, realized for instance with visible light in optomechanical gravitational wave detection, where they allow beating the shot-noise limit \cite{barsotti2018squeezed}. In the microwave regime, propagating squeezed microwaves are an indispensable ingredient of emerging quantum communication and quantum sensing technologies \cite{casariego2022}: protocols for quantum teleportation \cite{di2015quantum,fedorov2021experimental,gonzalez2022open} and for quantum key distribution \cite{fesquet2024demonstration,fesquet2023perspectives} have been realized in cryogenic environments, while open-air extensions are investigated. Quantum enhanced interferometry can beat the standard quantum limit \cite{kronowetter2023quantum} and quantum illumination \cite{lloyd2008enhanced,tan2008quantum,Karsa_2024} can offer a quantum advantage for imaging and radar applications under certain conditions \cite{assouly2023quantum,chang2019,messaoudi2020practical,barzanjeh2020microwave,Peichl20024}. Less explored is the use of sources of squeezed states in quantum information technologies, where they could find use for autonomous remote entanglement stabilization \cite{Didier2018}.

The eponymous characteristic of squeezed light is, that fluctuations are reduced beyond the vacuum level for one generalized quadrature with respectively enhanced fluctuations in the other. Clearly this distinguishes one distinct fixed direction in a (rotating-frame) phase space. A source, which keeps this direction stable, is thus crucial. While in ac-driven sources, such as various types of Josephson parametric amplifiers \cite{Eichler2011,Bergeal2012,flurin2012generating,zhong2013squeezing,svensson2018period,gu2017,wustmann2019parametric} the distinguished phase-space direction is simply determined by the phase of the ac-drive, dc-driven devices lack such a reference and consequently suffer from phase diffusion. Here, we investigate how stabilization can nonetheless be achieved by locking onto a weak oscillating reference signal.

We do this for a specific recently developed platform 
\cite{Hofheinz2011,Chen2014,Cassidy2017,Gramich2013,Armour2013}
for creating squeezed microwave light 
\cite{Westig2017,Peugeot2021, Padurariu2012,Juha2013,Armour2015,Trif2015}
by a \emph{dc-biased} Josephson junction connected in series to microwave resonators. One appeal of such \emph{Josephson-photonics devices} is the versatility of these sources, as a variety of different interesting quantum states \cite{Rolland2019,Grimm2019,Menard2022} can be created by a simple change of the dc-bias.

However, being dc-biased they exhibit a phase instability directly linked to fluctuations of the voltage dropping across the junctions, as $\varphi \sim \int dt\, \delta V(t)$ \cite{Wang2017}. 


A model of these devices and their phase diffusion caused by shot-noise of the Josephson current and an investigation of locking at the fundamental resonance, where each tunneling Cooper pair creates one photonic excitation, has been presented elsewhere \cite{hoehe2023quantum}. Here, we expand this description to the locking of squeezed states, encountered when each Cooper pair creates two photons. 
Besides being of immediate technological relevance, the locking of squeezed states shows strikingly different novel features. In contrast to a standard locking scenario the phase stabilization of a squeezed state relies on the injection of a signal at twice the frequency of the emitted radiation, which we want to lock. The way in which the radiation spectrum is modified by locking will thus obviously differ from the conventional case, where locking frequency and frequency to-be-locked (nearly) coincide \cite{Pikovsky_Rosenblum_Kurths_2001}.
This generic feature of locking of squeezed states has to our knowledge not been investigated before.
We will first consider Josephson-photonics devices operating in a regime, where they approximately realize generic single-mode and two-mode squeezing Hamiltonians and study the locking of the corresponding squeezed states by an injected ac-voltage signal, before considering the peculiar strong quantum nonlinearities, these devices can also exhibit, and their influence on locking. Finally, we will briefly consider the symmetry-breaking case of injecting a near-resonant microwave signal directly into the cavity, which resembles experiments performed in ac-driven systems \cite{Bengtson2018,markovic2019injection}.

\section{DC-generated single-mode squeezing}
\label{sec:sec_singlemode}

\begin{figure*}[t]
    \centering
    \includegraphics[width=1\linewidth]{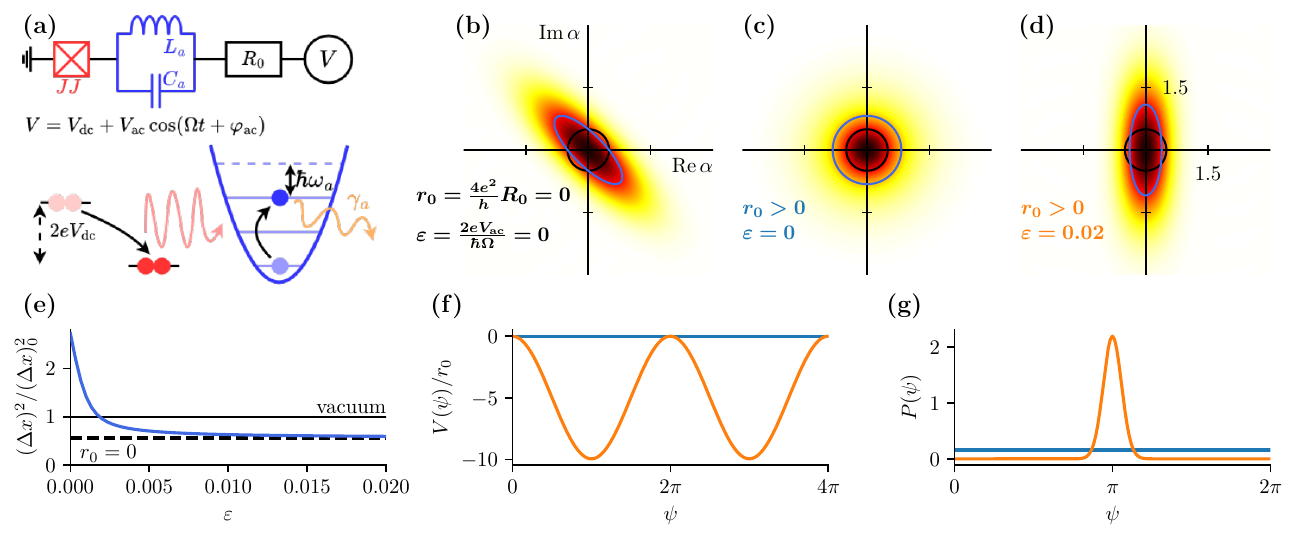}
    \caption{Single-mode squeezing. 
    (a) Circuit model of a Josephson-photonics device, where an $LC$-resonator (with eigenfrequency $\omega_a$ and single-photon loss rate $\gamma_a$ is connected in series with a Josephson junction. The circuit is resonantly dc-biased, $2eV_\mathrm{dc} \approx 2\omega_a$, such that each tunneling Cooper pair creates two photons in the resonator. (b) An idealized device with a fluctuationless dc-bias will reach a steady state which is squeezed [blue ellipse and dashed line in (e)] below vacuum level [black circle and gray line in (e)]. The ellipse angle is constant in the frame rotating with $2eV_\mathrm{dc}/2$ and set by the driving phase. (c) Voltage fluctuations through Cooper-pair shot noise, modeled by a small in-series resistance $R_0$ lead to diffusion of the driving phase, effectively resulting in the loss of squeezing and the $\mathbb{Z}^{k=2}$-symmetry. The ideal picture can be recovered (d) by adding an ac-locking signal with frequency $\Omega \approx 2eV_\mathrm{dc}/\hbar$ of sufficient amplitude $\varepsilon$ [blue curve in (e)]. The locking signal effectively creates a potential $V(\psi)$ for the driving phase with minima (f), in which it can be trapped. A competition between the force $\dot{\psi}=-V'(\psi)$ and noise yields (g) a steady-state probability distribution $P(\psi)$ centered around the potential minimum.
    [Parameters: $E_J^* = 80, \alpha_0 = 0.1, \delta_\mathrm{dc}=0, r_0  = 1/500$. The ellipse's minor and major axes indicate standard deviations of dimensionless phase space quadratures in the displayed state, whereas the black circle's radius indicates the same for the vacuum state.]}
    \label{fig:singlemode_wigner}
\end{figure*}

The `Josephson-photonics' platform we consider here for creating microwave with non-classical nature, such as various squeezed states, consists of a dc-voltage biased Josephson junction connected in series to an LC-stripline resonator and a resistance $R_0$, see figure~\ref{fig:singlemode_wigner}(a).

The system is described by a Hamiltonian 
\begin{equation}
    \label{eq:basic_josephson}
    \hat{H} = \hbar \omega_a \hat{a}^\dag \hat{a} - E_J\cos\left[\omega_\mathrm{dc}t - \varphi_{R_0} +\alpha_0( \hat{a}^\dag + \hat{a}) \right] \, 
\end{equation}  
amended by a Lindblad operator $\hat{L}_0 =\sqrt{\gamma_a}\hat{a}$ \cite{Gramich2013,Armour2013}.
The resonator hosts a single mode of frequency $\omega_a$ with an (external) quality factor $\omega_a/\gamma_a$ and a phase variable, $\alpha_0 ( \hat{a}^\dag + \hat{a})$, with zero-point fluctuations $\alpha_0 = (2e^2\sqrt{L/C}/\hbar)^{1/2}$. Besides the resonator's energy, the Hamiltonian contains a Josephson term, where the phase across the junction, appearing as the argument of the cosine, is found by Kirchhoff's rule adding up the phases across elements to find the phase, $\omega_\mathrm{dc} t = \int dt \; 2e V_\mathrm{dc}/\hbar$, associated with the bias voltage. Crucially, the model includes the integrated voltage dropping across the (small) resistance $R_0$ if a current $I_\mathrm{CP}$ flows,   
\begin{equation}
 \frac{\hbar}{2e} \varphi_{R_0}(t) =  \int_0^t d\tau\, V_{R_0}(\tau)  =  R_0 \int_0^t d\tau\, I_\mathrm{CP} =  R_0  2e \  m_\mathrm{CP}(t)\,,
 \label{eq:phase}
 \end{equation} 
which is connected to the number of Cooper pairs $m_\mathrm{CP}$ that have crossed the junction.
A dc current across the junction arises as Cooper pairs tunnel across the biased junction and lose energy by creating photonic excitations in the resonator, from which they eventually leak out via a transmission line. Clearly, we may expect various resonances when tuning the dc-voltage such that $\omega_\mathrm{dc} \approx k \omega_a (k \in \mathbb{N})$ and $k$ photons are created. Considering $k=2$, we find that for small detuning $\delta_\mathrm{dc} = \omega_a - \omega_\mathrm{dc}/2  \ll \omega_a$ and in the limit of small zero-point fluctuations, see \ref{sec:ap_rwa_sm_sympres}, the effective Hamiltonian in the rotating frame (defined by $\hat{U}(t) = e^{i \hat{a}^\dag \hat{a} \omega_\mathrm{dc} t/2}$) takes the form 
\begin{equation}
  \hat{H}_\mathrm{rwa} \approx \hbar \delta_\mathrm{dc} \hat{a}^\dag \hat{a} + \frac{E_J^* \alpha_0^2}{4} \cdot (e^{i\varphi_{R_0}}\hat{a}^{\dag 2} + e^{-i\varphi_{R_0}}\hat{a}^2) 
  \label{eq:ham_sm_rwa_loworder}
\end{equation}
of a squeezing Hamiltonian with $\varphi_{R_0}$ appearing as the phase of the squeezing parameter. Without a resistance, $R_0=0=\varphi_{R_0}$, one finds, indeed, a squeezed state, see figure~\ref{fig:singlemode_wigner}(b) for the steady-state Wigner function, where the $Z^{k=2}$ symmetry in phase-space reflects the symmetry of the Hamiltonian. 
Whereas the state has a trivial dependence on $\varphi_{R_0}$ (i.e., the squeezing angle is $\varphi_{R_0}/2 - \pi/4$), many observables such as occupation $\hat{n}$ and current $\hat{I}_\mathrm{CP}$ have such symmetry that their expectation values are independent of $\varphi_{R_0}$. 
The degree of squeezing increases with the renormalized Josephson energy $E_J^* = E_J \cdot e^{-\alpha_0^2/2}$ and the maximum suppression of one quadrature fluctuation below the vacuum fluctuations, the so-called of $3 \mathrm{dB}$ limit, is approached for $\alpha_0 \ll 1$ when driving at  threshold $E_J^*\alpha_0^2 \rightarrow \hbar\gamma_a$ \cite{Armour2013}
[A divergence, which appears for $\alpha_0 \rightarrow 0$, respectively for the Hamiltonian (\ref{eq:p2approx}), in photon number and anti-squeezed quadrature fluctuations is regularized for finite $\alpha_0$ by higher-order correction terms neglected in equation~(\ref{eq:p2approx}) but kept in 
figure~\ref{fig:singlemode_wigner}.].

\subsection{Phase diffusion and locking}

Including now a finite resistance, $R_0 \neq 0 $, the phase $\varphi_{R_0}$ no longer remains fixed to zero, but increases over time with each Cooper pair tunneling across the junction. In fact, there will be a mean growth determined by the mean Cooper-pair current and fluctuations caused by its shot-noise. The consequent diffusion of the phase $\varphi_{R_0}$ leads to an 
effective averaging over all squeezing angles as the phase variance grows beyond $2\pi$ and the steady state under phase-diffusion is rotationally symmetric [cf. figure~\ref{fig:singlemode_wigner}(c)]. 
No squeezing is observable, and, in fact, fluctuations in any direction are larger than vacuum fluctuations. 
To stabilize the squeezing ellipse we add a small
 ac-voltage, $V(t) = V_\mathrm{dc} + V_\mathrm{ac} \cdot \sin(\Omega t + \varphi_\mathrm{ac})$ as a locking signal which breaks rotational symmetry but does not break the $\mathbb{Z}^{k=2}$ symmetry, which is the case for  $\Omega \approx  2 \omega_a \approx \omega_\mathrm{dc}$.
As expected, a sufficiently strong locking signal restores the undiffused phase-space squeezing ellipse as stationary state of the system in the frame rotating with $\Omega/2$, see figure~\ref{fig:singlemode_wigner}(d), and recovers nearly the full sub-vacuum noise squeezing of the $R_0=0$ case, see figure~\ref{fig:singlemode_wigner}(e).

For the simulations of phase diffusion and locking as shown in figure~\ref{fig:singlemode_wigner}(c, d) and for an understanding of the locking mechanism we have to employ a model that goes beyond simple Lindbladian dynamics with the Hamiltonians (\ref{eq:ham_sm_rwa_loworder}) and (\ref{eq:singlemode_rwa_rot}) with a fixed phase $\varphi_{R_0}$, but account for the feedback of the stochastic Cooper pair current on that phase, as described by equation~(\ref{eq:phase}). Such a model, relying on an extension of the number-resolved master equation technique \cite{Xu2013} to the counting of the number $m_\mathrm{CP}$ of Cooper-pairs transported in a \emph{coherent} tunneling process, has been established in \cite{hoehe2023quantum} for a Josephson-photonics device operated at a voltage bias, where each Cooper pair excites a single photon and the oscillator is driven into a coherent state for small zero-point fluctuations. Locking is then achieved by including the ac-voltage in the Hamiltonian (\ref{eq:basic_josephson}), $\omega_\mathrm{dc} t \rightarrow \int_0^t d\tau\, V(\tau)$, which in the frame rotating with $\Omega/2$ can be approximated (see \ref{sec:ap_rwa_sm_sympres}) in lowest order of zero-point fluctuations and locking strength $\varepsilon=2e V_\mathrm{ac}/\hbar\Omega$ 
\begin{equation}
    \label{eq:p2approx}
    \hat{H} \approx \left[\hbar (\delta_\mathrm{dc} +\delta_\mathrm{ac}) - 
    \varepsilon \frac{E_J^*\alpha_0^2}{2} \cos\psi \right] 
    \hat{a}^\dag \hat{a} +\frac{E_J^* \alpha_0^2}{4} \cdot (e^{i\psi}\hat{a}^{\dag 2} + e^{-i\psi}\hat{a}^2)
\end{equation}
where $2\delta_\mathrm{ac}=\omega_\mathrm{dc} - \Omega$ is the detuning between the frequencies of the ac- and the dc-drive and $\psi = (\Omega t + \varphi_\mathrm{ac}) -(\omega_\mathrm{dc} t - \varphi_{R_0})$ is the phase difference between the ac-driving phase and the dc-driving phase. 
The ac-signal yields an additional detuning term in the Hamiltonian which depends on the phase difference $\psi$. This term stems from tunneling processes, which are not changing the occupation of the cavity: the absorption of an ac-drive photon drives a Cooper pair tunneling against the dc bias or in the reverse process a Cooper pair tunnels with the bias and emits a photon into the drive. These processes, which are proportional to $\varepsilon E_J$, are renormalized by the presence of the cavity and the concomitant (virtual) emission and re-absorption of cavity photons. While the term in the Hamiltonian, which is the lowest-order expansion of a normal ordered Bessel-function $J_0(2 \alpha_0 \sqrt{\hat{a}^\dag \hat{a}})$, cf. equation~(\ref{eq:singlemode_rwa_rot_hdag}) in \ref{sec:ap_rwa_sm_sympres}, reflects the minor (back-action) effect of this tunneling processes on the cavity, the same processes will also contribute to the current:
\begin{equation}
  \label{eq:current_sm_rwa_loworder}
    \frac{\hat{I}_{CP}}{2e} \approx \frac{E_J^* \epsilon}{2\hbar} \sin(\psi)  \left(1 - \alpha_0^2 \hat{a}^\dagger \hat{a}  \right) 
  + i \frac{E_J^*\alpha_0^2}{4\hbar} \left(e^{-i \psi} \hat{a}^2 -  e^{+i \psi}\hat{a}^{\dagger 2} \right)  \,.
\end{equation}
The second term with photon pair creation and annihilation operators is the conventional Josephson-photonics current for a pure dc-bias, which results in a $\psi$-independent current for $\varepsilon=0$. The first term is the extra contribution added by the ac signal and hence proportional to $\varepsilon$. That current arises as the difference of the aforementioned forward and backward processes and consequently yields the sine term, which now also includes an ${\cal{O}}(\alpha_0^0)={\cal{O}}(1)$ term without virtual cavity photons that is irrelevant in the Hamiltonian. It is this Shapiro-like term \cite{Shapiro1963}, which can lock the system. As $\dot{\psi}= -2\delta_\mathrm{ac} + \dot{\varphi}_{R_0}$ and $\dot{\varphi}_{R_0}\propto I_{CP}(\psi)$ instead of $\psi$ being a fixed parameter its dynamic is governed by a stochastic differential equation. Average drift and fluctuations of $\psi$ depend on mean and fluctuations of the CP current which itself depends on $ \psi$. The resulting reduced dynamics can be understood as diffusive (overdamped) motion in a potential \cite{Kramers1940,risken1996fokker}, whose form can be formally derived via two-time perturbation theory, see \ref{sec:ap_tt} and \cite{DoMu2018,pavliotis2008multiscale,hoehe2023quantum}.

Without locking drive the current is independent of $\psi$, so that we get diffusion on a potential $V(\psi)\propto \psi$ with a slope related to the detuning including an average voltage drop at $R_0$. The phase $\psi$ will therefore explore its full  parameter space, leading to the rotationally symmetric Wigner density of figure~\ref{fig:singlemode_wigner}(c) without squeezing.
With locking drive, the expectation value of the current oscillates with $\psi$ around a mean value, eventually allowing for the creation of maxima and minima in the tilted Adler-type potential \cite{Adler1964}, see figure~\ref{fig:singlemode_wigner}(f). The phase-particle can become trapped in one of the minima and the phase is constant but for small variations. As demonstrated in figure~\ref{fig:singlemode_wigner}(d) this restores the undiffused squeezing ellipse of figure~\ref{fig:singlemode_wigner}(b) with some minute additional broadening due to the remaining fluctuations of $\psi$. 

While diffused and deeply locked state are clearly distinguishable, phase-space properties are typically not the observables most suited to clearly indicate a locking transition. This can be seen in the very smooth crossover of the quadrature with increasing the locking strength in figure~\ref{fig:singlemode_wigner}(d) and can be understood from the dynamics in the locking potential: values of $\psi$, where the slope is reduced (increased), will be more (less) likely than the average, well before the emergence of local minima at larger locking signal $\varepsilon$.\\

\subsection{Correlation functions}
\begin{figure*}[b]
    \centering
    \includegraphics[width=1\linewidth]{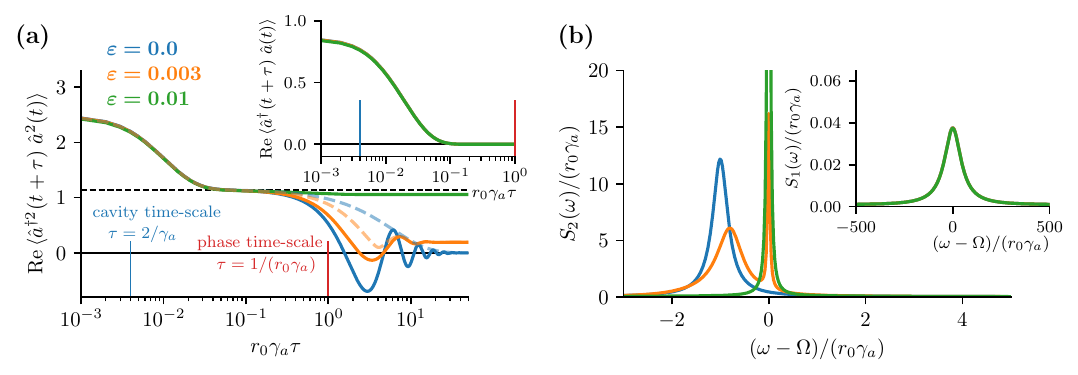}
    \caption{First-order correlation functions $G^{(1)}$ (insets) and the unconventional second-order correlation functions $G^{(2)}$ in the time domain (a) and the corresponding frequency dependent spectra (b). The conventional spectrum [inset of (b)] of squeezed light is broadened by the cavity decay rate for weak driving, but becomes substantially sharpened when driven closer to the squeezing divergence as done in (b). In time domain it decays towards $|\langle \hat{a} \rangle|^2=0$ for times larger than the timescale of the cavity dynamics irrespective of the much slower dynamics of the cavity phase. 
    The (unconventional) second-order correlation function $G^\phi$ instead shows cavity and phase dynamics time scales (a). After an initial partial decay to $|\langle \hat{a}^2 \rangle_{r_0=0}|^2$ (dashed line) on cavity time scales, which is independent of the ac-signal strength, the phase diffusion determines the further behavior. In frequency domain (b) the emergence of a dominant spectrally sharp contribution at $\Omega$ corresponding to a finite $t\rightarrow \infty$ limit indicates locking. Width and position of the broad peak reflect the corresponding time evolution on the time scales of the slow phase dynamics and yield direct insight into the phase-phase correlations of the diffusive motion in the potential, cf. figure \ref{fig:singlemode_wigner}~(f).
    [Parameters: $E_J^*=80, \alpha_0=0.1, \delta_\mathrm{dc}=0, \nu_0=1 r_0 \gamma_a, r_0=1/500$. Delta peaks are artificially broadened by $\gamma_a/1000$ (FWHM).]}
    \label{fig:singlemode_spectrum}
\end{figure*}

For the locking of a conventional (not squeezed) oscillator an experimentally readily observable hallmark of a locked state is the sharpening of the resonators noise-broadened emission spectrum \cite{Pikovsky_Rosenblum_Kurths_2001,Cassidy2017}. However, squeezing is different: as each tunneling Cooper pair creates two photons and distributes its energy between the two, only the sum of the photons frequency is fixed by  $\omega_1 + \omega_2 = \omega_\mathrm{dc}$, while each individual photons frequency and thus the spectrum is lifetime broadened even without voltage fluctuations. These will only add insubstantially to the broadening (for typical experimental parameters) and locking can at best curtail this small extra broadening, see figure~\ref{fig:singlemode_spectrum}(a).
The closest analogue to the locking-induced spectral sharpening is found in a different experimentally accessible observable that can reveal phase-phase correlations,
\begin{equation}
\label{eq:g_phi}
G^\phi(t+\tau, t) = \braket{\hat{a}^{\dag 2}(t+\tau) \hat{a}^2(t)} \,.  
\end{equation}
In a recent Josephson-photonics experiment, such observables were used as entanglement witness for two-mode squeezed radiation, cf.~\cite{Peugeot2021}, and section~\ref{sec:sec_twomode} below.
Similar to the conventional (unnormalized) second-order correlation function $G^{(2)}(\tau)$, on very short time scales  ($\gamma_a \tau \lesssim 1$) the observable $G^\phi(\tau)$ reflects the bunching of photons due to the pair creation process and the relaxation dynamics of such excitations \cite{dambach2015time}. For times much longer than all internal  cavity relaxation and excitation dynamics, the correlation function factorizes,  $G^\phi(\tau) \rightarrow \langle (a^\dagger)^2(t+\tau) \rangle  \langle a^2(t) \rangle$, and only the phases of the two terms remain correlated. Without fluctuations and locking signal, these phases strictly oscillate with the applied dc-voltage, $G^\phi(t+\tau, t)\rightarrow |\braket{\hat{a}^2}_\mathrm{st}|^2\exp{[-i \omega_\mathrm{dc} (t+\tau-t)]}$, and $g^\phi(\tau)$ becomes constant in the corresponding rotating frame.  Phase-diffusion is thus directly reflected in the long-time-decay of $g^\phi(\tau)$ shown in the blue line of figure~\ref{fig:singlemode_spectrum}(b). In our model shot-noise yields an exponential decay of phase-phase correlations on the scale $D \tau \sim 1$, where the free diffusion parameter $D(\psi) \approx D = 0.164 r_0 \gamma_a$ can be calculated via two-time perturbation theory \cite{hoehe2023quantum,DoMu2018,pavliotis2008multiscale}, and a Lorentzian spectral broadening in Fourier-space. In the frame rotating with the locking signal $\Omega$ the decay is overlaid by oscillation with the detuning $\nu_0$ of the broad unlocked peak from $\Omega$. Note that in recent experiments \cite{Peugeot2021,Menard2022} quasi-static classical fluctuations instead yielded Gaussian shapes in time and frequency. Adding a sufficiently strong locking signal, however, will constrain phase-diffusion in the minima of the potential and preserve some phase-phase correlations, such that $G_{\phi,\infty}$ remains nonzero, leading to a delta-peaked spectrum (green line in figure~\ref{fig:singlemode_spectrum}) while the remaining broad part in the Fourier spectrum and the corresponding partial decay is caused by diffusion within one minimum. 

Before revisiting the single-mode case to investigate few-photon nonlinearities and symmetry-breaking locking by a directly injected ac-signal at a different frequency in sections~\ref{sec:sec_singlemode_ZPF} and \ref{sec:sec_singlemode_sbl} below, we now turn to the case, where each Cooper pair creates photons in two different modes.

\section{Two-mode squeezing}
\label{sec:sec_twomode}

Biasing at a resonance, $2 e V_\mathrm{dc}=\hbar (\omega_a +\omega_b)$, a Josephson-photonics device can create two-mode squeezed states \cite{Westig2017,Peugeot2021,Armour2015,Trif2015}. The two modes can be realized within a single resonator or in two cavities connected in series by the junction, which allows for an easy separation of the emission from the cavities into two distinct microwave striplines.
A bright and simple quantum source of two-mode squeezed microwave light is of high technological relevance, as it enables applications such as (continuous variable) quantum teleportation \cite{di2015quantum,fedorov2021experimental} and quantum illumination protocols proposed as a potential quantum radar \cite{Karsa_2024,assouly2023quantum,chang2019,messaoudi2020practical,barzanjeh2020microwave} and many more as explained in the introduction.

\begin{figure*}[b]
    \centering
    \includegraphics[width=1\linewidth]{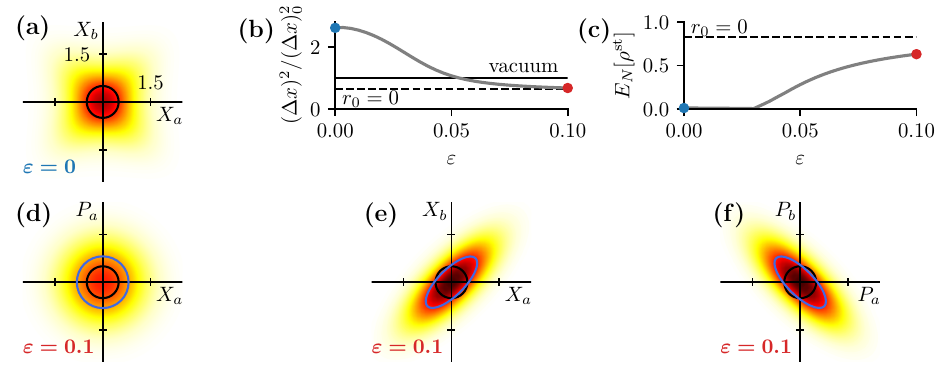}
    \caption{
        Phase locking of two-mode squeezed light. (a) Mixed phase space without locking. (c, d) In mixed phase spaces $(X_a, P_a)$ and $(X_b, P_b)$  the injected ac-signal restores the phase coherence of the squeezing ellipse. (b) Still, the coherence vanishes in the reduced phase space of each cavity. (e) The width $\Delta x$ of the squeezing ellipse decreases below zero-point fluctuations of the vacuum, remaining ultimately limited to 3 dB. (f) Entanglement of photons characterized by the logarithmic negativity $E_N$ is restored.
        [Parameters: $E_J^*=80, \alpha_0=\beta_0=0.1, \delta_\mathrm{dc}=0, \nu_0=0, r_0=1/700$. Rotating frame angles for $a$ and $b$ are chosen, so that maximal squeezing appears in the $X_a-X_b$ and $P_a-P_b$ reduced phase spaces.]}
    \label{fig:twomode_tt}
\end{figure*}

Transforming the two-mode generalization of the Hamiltonian, equation~(\ref{eq:basic_josephson}), 
\begin{equation}
\label{eq:Hamiltonian_twomode_org}
\hat{H} =\hbar \omega_a \adag \ann + \hbar \omega_b \bdag \bnn - E_J\cos\left[ \omega_\mathrm{dc}t + \alpha_0 (\adag + \ann) + \beta_0 (\bdag + \bnn) - \varphi_{R_0}\right]     
\end{equation}
to a frame rotating with frequency $\Omega$, the rwa-Hamiltonian,
\begin{equation}
\label{eq:hamiltonian_twomode}
    \hat{H}_\mathrm{rwa} = \hbar\delta\frac{ \hat{a}^\dag \hat{a} + \hat{b}^\dag \hat{b}}{2} 
        + \frac{E_J^*\alpha_0\beta_0}{2}(e^{i\psi}\hat{a}^\dag \hat{b}^\dag + e^{-i\psi}\hat{a}\hat{b})
\end{equation}
where the detuning is $\delta = \omega_a + \omega_b - \Omega =  \delta_\mathrm{dc} + \delta_\mathrm{ac}$ and $E_J^* = E_J e^{-(\alpha_0^2 + \beta_0^2)/2}$. Detunings, $\delta_\mathrm{dc} = \omega_a + \omega_b - \omega_\mathrm{dc}$ and $\delta_\mathrm{ac} = \omega_\mathrm{dc} - \Omega$, are defined analogously to the single-mode case, as is the phase drop across an in-series resistor, $\varphi_{R_0}$, and its relation to the Cooper pair current.

Again in equation~(\ref{eq:hamiltonian_twomode}) we only kept those term of lowest order in the zero-point fluctuations $\alpha_0, \beta_0$ relevant for the squeezing and locking physics (see e.g. \cite{Armour2015,Trif2015} and the \ref{sec:ap_rwa_tm_sympres} for the full Hamiltonian used also for all numerical results below).

Adding a weak ac-locking signal with a frequency $\Omega$ close to $\omega_\mathrm{dc}$ again adds a detuning term to the Hamiltonian
\begin{equation}
        \hat{H} = \hat{H}_\mathrm{rwa} - \varepsilon \frac{E_J^*}{2}\cos\psi \cdot (\alpha_0^2 \adag \ann + \beta_0^2 \bdag \bnn) \,.
\end{equation}

For two-mode squeezing the applied voltage determines the phase of the common bilinear drive term, $a^\dagger b^\dagger$, corresponding to a sum of phase-space angles for the $a$ and $b$ cavities. Including the resistance and the locking signal, it will be this combined phase, which acquires a stochastic dynamics in an effective potential $V(\psi)$, which will diffuse, and which eventually can be locked. For the phase-space distributions, $W(X_a,P_a, X_b, P_b)$, this means, that the features of an undiffused two-mode squeezed state \cite{Eichler2011,fedorov2018finite}, which are destroyed by voltage fluctuations, see figure~\ref{fig:twomode_tt}(a), can be restored, when locking the system to a sufficiently strong ac-signal, see dashed lines in figure~\ref{fig:twomode_tt}(e,\,f). Figure \ref{fig:twomode_tt}(b) shows the rotationally symmetric, thermal Wigner distribution of the reduced density matrices for cavity $a$ (and similar for $b$), and the squeezing ellipses in a combined phase-space of $a$ and $b$ quadratures 
figure~\ref{fig:twomode_tt}(c),\,(d). Entanglement measured by the logarithmic negativity is also restored by locking, although remaining diffusion around the locking minimum of $V(\psi)$ will lead to some degradation, cf. dashed black versus blue line in figure~\ref{fig:twomode_tt}(f). The proposed locking mechanism thus overcomes a crucial impediment for employing a quantum source lacking phase stability, an inherent drawback of dc-driven devices, for quantum information transfer and processing technologies. 

\section{Beyond the squeezing limit}
\label{sec:sec_singlemode_ZPF}
\begin{figure*}[b]
    \centering
    \includegraphics[width=1\linewidth]{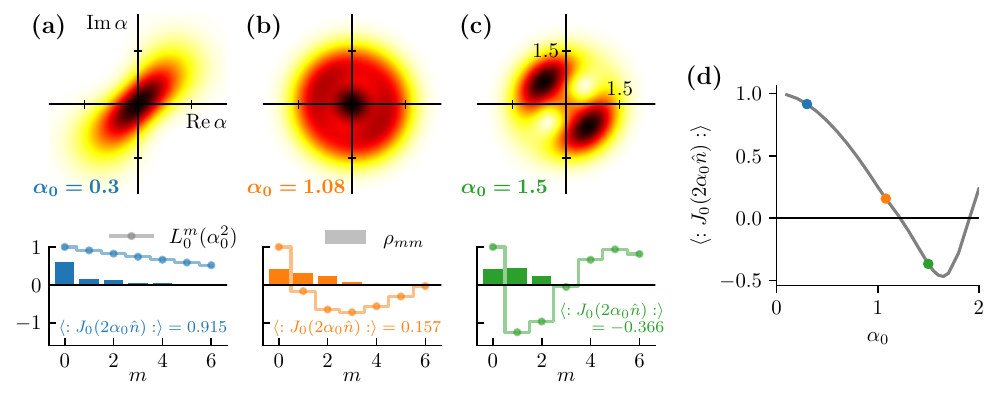}
    \caption{Beyond the squeezing limit of weak driving and low zero point fluctuations $\alpha_0$ a nontrivial dependence of the locked Wigner distributions is found upon increasing $\alpha_0$ is found. This is mainly due to a renormalization of the Shapiro-like effective locking term by a Bessel functions $J_0(2\alpha_0 \hat{n})$ describing virtual emission and absorption of cavity photons. For moderate zero point fluctuations (a) the matrix elements of the normal-ordered Bessel functions $L_0^m \sim 1$ for all occupied states and the (slightly deformed) squeezing ellipse is well locked. Increasing $\alpha_0$ a region without locking potential (b), and (c) an eventual reversal of the potential resulting in a state locked perpendicular to the original orientation is found. The vanishing of the locking potential in (b) does not exactly coincide with the zero of $\langle J_0\rangle$ due to higher-order correction terms. [Parameters: $\braket{n} = 1, \delta_\mathrm{dc} = 0, r_0 = 1/1000, \varepsilon=0.01$.]
    }
    \label{fig:beyond_squeezing}
\end{figure*}

So far, we have considered the locking of one- and two-mode squeezed states created by a Josephson-photonics device operated in the bilinear regime of the approximated Hamiltonian~(\ref{eq:p2approx}) and (\ref{eq:hamiltonian_twomode}). These are approximations of more complex normal-ordered Bessel-function terms, e.g.
$\ann^2 \rightarrow :2\ann^2 J_2(2\alpha_0 \sqrt{\hat{n}})  / (\alpha_0\sqrt{\hat{n}})^2:$, which are valid if the arguments of the Bessel function, $2\alpha_0 \sqrt{\hat{n}} \ll 1$, where higher Fock states $| n \rangle$ become relevant, when the effective driving strength, $E_J^* \alpha_0^2$ ($E_J^* \alpha_0 \beta_0$ for two modes), is increased.
Turning now, therefore, to the regime of strongly nonlinear driving of the single-mode device, in particular to larger zero-point fluctuations $\alpha_0$, so that nonlinearities are becoming pertinent on the few-photon level, the Josephson-photonics dynamics and the locking features change substantially. 
We first briefly recapitulate the well-studied \cite{Kubala2015,Kubala2020,lang2021,Menard2022} effects on the system without resistance and locking signal: In the Hamiltonian the squeezing terms are amended by higher-order nonlinear corrections terms diagonal in the number basis, which can conveniently be expressed as a series expansion of the normal-ordered Bessel function or as transition matrix elements involving Laguerre polynomials, $\langle n+2 | H |n\rangle \propto L_{n}^{(2)}(\alpha_0^2)$ (see\,\ref{sec:ap_laguerre_mat_elem}). 
While for the pure squeezing Hamiltonian (\ref{eq:p2approx}) occupation and major axis of the squeezing ellipse diverge for $E_J^*\alpha_0^2/(\hbar\gamma_a) \rightarrow 1$, nonlinearites of the full Hamiltonian limit this growth and as the effective driving increases  across the erstwhile divergence while $\alpha_0 \ll 1$ the squeezing ellipse splits into two parts, cf. figure~\ref{fig:beyond_squeezing}(c), which eventually branch out along a circle \cite{lang2021,Menard2022}. Larger $\alpha_0$ leads to a smoothened crossover and stronger deformations of the ellipsoid.

Adding the ac locking signal to the dc bias, the Hamiltonian, equation~(\ref{eq:ham_sm_rwa_loworder}), and the current, equation~(\ref{eq:current_sm_rwa_loworder}), are furthermore modified in similar fashion: most crucially the Bessel-function $J_0$ appears in the $\cos \psi$ term in the Hamiltonian and the corresponding $\sin \psi$ term of the current, which describe as explained above the renormalization effects of exchange of photons with the cavity on the $E_J \varepsilon$ processes of ac-assisted Cooper pair tunneling. In addition to this modification of the term solely responsible for the locking at $\alpha_0 \ll 1$, there appears an additional locking-enabling term $\propto E_J \varepsilon \alpha_0^4 \hat{a}^4 + c.\,c.$ creating higher-order squeezing.
To illustrate how the renormalization terms can dramatically modify the locking features, we consider in figure~\ref{fig:beyond_squeezing} three different values of zero point fluctuations and chose corresponding driving strengths, so that the average cavity occupation $\langle \hat{n} \rangle = 1.0$ (for $r_0=0$).
Increasing the zero point fluctuations, we find a transition from locked to unlocked and back to locked with the squeezing axis rotated by $\pi/2$. While this reflects the behaviour of the Bessel-function $J_0(x)$ moving through its first zero at $x \sim 2 \alpha_0 \langle \hat{n} \rangle$, for the squeezed state and even more so for the strongly nonlinear squeezing for $\alpha_0 \sim 1$ instead of the mean occupation $\langle \hat{n} \rangle$ the occupation distribution $P(n)=\langle n|\hat{\rho}|n\rangle$ has to be considered, see bars in figure~\ref{fig:beyond_squeezing}. Weighting these occupations with the matrix elements $\langle n|e^{i \alpha_0 (\hat{a}^\dagger + \hat{a})}|n\rangle$ corresponding to $J_0$ yields the effective strength and sign of the main locking term, see lines and numerical values in figure~\ref{fig:beyond_squeezing}. The newly appearing higher-order squeezing term also contributes and shifts the de-locking transition to a slightly smaller $\alpha_0$ value.   

Staying with the single-mode case but reverting to smaller $\alpha_0$ and the squeezing approximation, we now turn to locking by injecting a microwave directly into the cavity instead of feeding it via the bias-line as above. 

\section{Symmetry breaking locking}
\label{sec:sec_singlemode_sbl}

Instead of adding a small ac-voltage to the dc-bias, locking can also be achieved by coupling an ac-oscillation directly to the resonator cavity, e.g., by sending a microwave signal into the striplines typically used as output port of the Josephson-photonics device. While in general, there are different coupling scenarios, such as inductive or capacitative corresponding to $\hat{x}_\mathrm{in}\cdot\hat{x}_a$ or  $\hat{p}_\mathrm{in}\cdot\hat{p}_a$ schemes, any weak bilinear coupling will ultimately add a linear driving term $\hat{a}_\mathrm{in}\cdot\hat{a}^\dagger + \mathrm{h.c.} \rightarrow \alpha_\mathrm{in}\cdot\hat{a}^\dagger + \mathrm{h.c.}$
after rotating-wave approximation and assuming a classical (coherent) drive.
As the ac-term no longer appears within the $\cos$-term of the Josephson-photonics Hamiltonian, higher powers of the oscillations and frequency mixing effects, which appeared above when feeding the ac-signal via the bias, are absent now. 
In consequence, the ac-locking signal is to be sent with a frequency $\Omega\approx \omega_a \approx \omega_\mathrm{dc}/2$ close to the cavity emission frequency we want to lock. That locking signal will, however, break the discrete time-translational invariance of the dc-drive at the $k=2$ parametric resonance, $\omega_\mathrm{dc} \approx k  \omega_a$ and the corresponding $\mathbb{Z}^{k=2}$ phase-space symmetry of the squeezing Hamiltonian. Nonetheless, one may aim for a locking signal amplitude small enough, so that this symmetry breaking is weak and a largely undisturbed squeezing ellipse with stable phase is restored.

\subsection{Hamiltonian and current}

Starting from the purely dc-driven ($\omega_\mathrm{dc}=2e V_\mathrm{dc}/\hbar$) Josephson-Photonics Hamiltonian (\ref{eq:basic_josephson}) with an additional drive term $\hat{H}_\mu= \varepsilon_\mu \cos{(\Omega t + \phi_\mu)}(\hat{a} +\hat{a}^\dagger)$ describing a direct microwave injection into the cavity, as in section~\ref{sec:sec_singlemode} and section~\ref{sec:sec_twomode} we can again consider the limit of small zero point fluctuations, so that the rotating wave Hamiltonian 
\begin{equation}
\label{eq:H_RWA_SBL}
\hat{H}_\mathrm{rwa} \approx \hbar \delta\adag \ann   + \frac{E_J^*\alpha_0^2}{4} \left(
e^{-i \psi} \hat{a}^2 + e^{+i \psi}\hat{a}^{\dagger2} \right) + \frac{\varepsilon_\mu}{2} \cdot \left(e^{i\phi_\mu} \adag + e^{-i\phi_\mu} \ann \right)
\end{equation}
results. The approximate Hamiltonian combines the squeezing term, $\propto E_J^*\alpha_0^2 \hat{a}^2 + c.c$, of the dc-drive with a linear term, $\propto \varepsilon_\mu \hat{a} + c.c$, from the direct ac-drive. Crucially, there are no Shapiro-like terms of $\propto \varepsilon_\mu^2 E_J^*$ form, where dc- and ac-drive exchange excitations 
and the dominant nonlinear corrections to the approximate Hamiltonian (\ref{eq:H_RWA_SBL}) will be the familiar Bessel-functions modifying the squeezing. Similar fundamental differences are found in the expression for the Cooper pair current, which is within rotating-wave approximation found as  
 \begin{equation} 
 \label{eq:I_CP_SBL}
  \frac{\hat{I}_\mathrm{CP}}{2e} 
  \approx\frac{E_J^*\alpha_0^2}{\hbar} \frac{1}{4} \left(i e^{-i\psi} \hat{a}^2 -i e^{+i\psi} \hat{a}^{\dagger 2} \right)\;.
\end{equation}
As in the Hamiltonian there are no Shapiro-like terms and only the dc-driven squeezing terms of the Hamiltonian appear, as only those terms involve Cooper pair tunneling.
Since it is the feedback of this Cooper pair current on the phase $\psi$
of the dc-driving terms which yields locking, not only the dynamics of the cavity state governed by the modified Hamiltonian (\ref{eq:H_RWA_SBL}) but also the fundamental locking mechanism determined by the current is fundamentally different from the afore-studied scenarios.

\begin{figure*}[b]
    \centering
    \includegraphics[width=\linewidth]{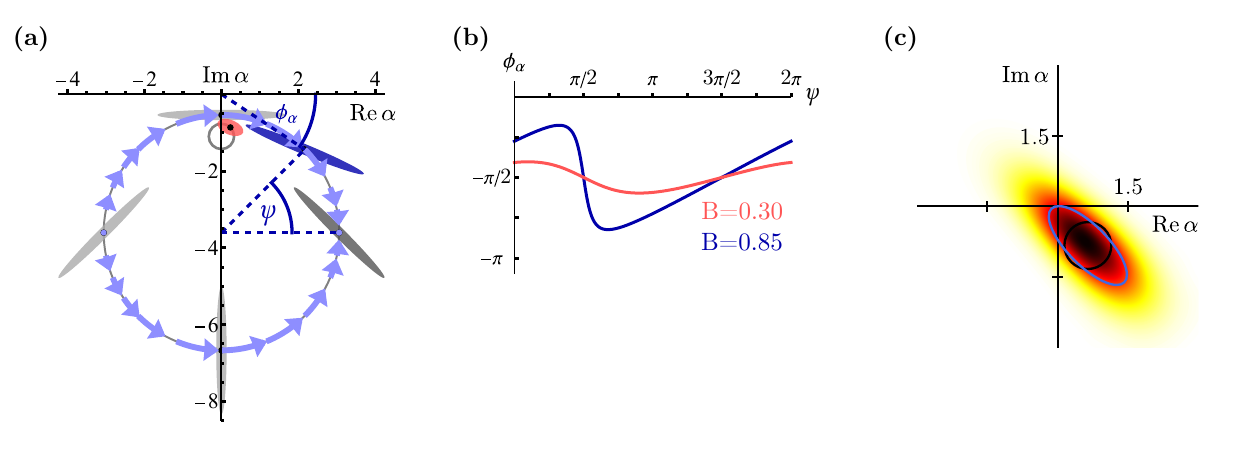}
    \caption{(a) Sketch of steady-state solutions of a bilinear Hamiltonian with effective squeezing strength $B=E_J^* \alpha_0^2/(\hbar \gamma_a)$ as the phase $\psi$ of the sqeezing is varied. The centers of the squeezing ellipses, described by the solutions for $\alpha=|\alpha|e^{i \phi_\alpha}$, are found on a shifted circle. The amount of the shift and the radius of the circle depend on the strengths of the microwave signal and of the squeezing drive (compare equation~(\ref{eq:ap_alpha_solution})). The major ellipse axis rotates with $\psi/2 -\pi/4$. For the large circle shown in blue, the strength and direction of the force $\dot{\psi}$ on the  squeezing ellipse caused by the current feedback on $\psi$ are indicated by blue arrows, resulting in the example in a locked solution  at $\psi=0$ (since $\omega_\mathrm{dc} = 2\omega_a + \frac{2e}{\hbar}R_0 \braket{\hat{I}_\mathrm{CP}(\psi=0)}_{\rho(\psi=0)}$ is chosen). (b) Solutions with a squeezing drive far below the divergence, $B=0.3 \ll1$ [cf. the red ellipse on the small circle in (a)], have their phase shift $\phi_\alpha$ undergoing small near-sinusoidal oscillations around $-\pi/2$, while solutions close to the divergence point with $B=0.85\approx 1$  [solutions on the large circle in (a)] have a strong sawtooth-like feature. (c) Wigner density of a squeezed state locked by a symmetry breaking, directly injected ac-signal. Shift and rotation are explained by the sketch in (a), while the squeezing below vacuum level (black circle) is indicated by the blue ellipse.  [Parameters for (a) and (b): $\epsilon_\mu = \hbar \gamma_a$, $\phi_\mu = 0$, $\Omega = \omega_a$. Parameters for (c): $E_J^* = 80, \alpha_0 = 0.1, \varepsilon=0.3 \hbar \gamma_a, r_0 = 1/2000.$]}
    \label{fig:symbreak_locking}
\end{figure*}

\subsection{Breaking the phase-space symmetry}

Still, we can proceed in the same manner as above, by first investigating the steady state of the cavity for a fixed $\psi$, which will then yield the current from which in turn the locking potential $V(\psi)$ follows.
The first step is, hence, an interesting though simple quantum optics problem of finding the steady state of a damped cavity driven in parallel by a squeezing term, resulting on its own in a squeezing ellipse centered around the origin, and a linear term, which by itself yields a displaced vacuum state. Pursuing the most elegant way to understand how these two driving terms coordinate their emission into a common  mode, we remove the linear drive by a shift of the cavity operators, $\hat{a}^{(\dagger)} \rightarrow \alpha^{(*)} + \hat{b}^{(\dagger)}$, with a properly chosen $\alpha$. 
Clearly, the strength of the squeezing term remains unchanged, so that the squeezing ellipse found in the absence of a linear term is merely displaced to a new position $\alpha$, which depends on strength and phase of the ac-drive, but the ellipse is neither shifted or rotated, nor is its form changed by the addition of that drive.
For the linear term of the Hamiltonian, we obtain, cf. equation~(\ref{eq:SMl_H_RWA_SBL_trafo}), 
\begin{equation}
 \label{eq:SBL_trafo}
      \bdag \left(\frac{\varepsilon_\mu}{2}e^{i \phi_\mu}  -i\hbar \frac{\alpha}{2} (\gamma_a + 2i\delta) + \alpha^* \frac{E_J^* \alpha_0^2}{2} e^{i \psi} \right) + \mathrm{h.c.} \;.
 \end{equation}
Finding the proper value of the shift $\alpha$ for which all linear contributions \---from performing the transformation in the linear and squeezing terms of the Hamiltonian and in the Lindbladian\--- cancel, can be mapped to a trigonometric exercise visualized in figure~\ref{fig:symbreak_locking}, where we consider the resonant case, $\delta=0$, for simplicity and set the phase of the ac-drive to zero, $\phi_\mu=0$. As shown in the \ref{sec:ap_symbreak_locking} the solutions for $\alpha=|\alpha| e^{i \phi_\alpha}$ describe a circle in the complex plane as $\psi$ is varied. Distinct dependencies of phase $\phi_\alpha$, see figure~\ref{fig:symbreak_locking}(b), and amplitude $|\alpha|$ of the shift on $\psi$ are found for squeezing far below and close to the squeezing divergence $E_J^*\alpha_0^2/(\hbar \gamma_a)\rightarrow 1$ (The divergence will, of course, again be regularized by higher-order terms in $\alpha_0$). These can be visually understood as resulting from the circle in figure~\ref{fig:symbreak_locking}(a) remaining far from or approaching the center of the complex plane, but may also easily be investigated analytically (see \ref{sec:ap_symbreak_locking}). 

Having gained a thorough understanding of the quantum optical interplay of linear and squeezing drive and the resulting stationary state $\rho(\psi)$ for a fixed phase $\psi$ of the dc-drive term \footnote{Notably, it is this interplay alone which leads to locking-type phenomena in recent experiments \cite{svensson2018period,markovic2019injection} on two-mode devices operated beyond the parametric threshold, which are ac-driven and, hence, do not exhibit dynamics of the driving phase and the feedback mechanism discussed here.}, we can now calculate the corresponding Cooper pair current (\ref{eq:I_CP_SBL}), which, while having no obvious counterpart in the generic quantum optics problem, is relevant to the phase dynamics of our problem. Somewhat surprisingly, much of the richness of different regimes and nontrivial behaviour of $\alpha(\psi)$ is lost in the current $\langle \hat{I}_\mathrm{CP}(\psi) \rangle_{\rho(\psi)}$, which shows simple sinusoidal oscillations scaling with $\varepsilon_\mu^2$ around an average. From figure~\ref{fig:symbreak_locking}(a) one can easily guess that (for $\delta=0=\phi_\mu$) the current is maximal, when the shift of the squeezing ellipse is along its major axis as occurs at the south pole of the circle, and minimal on the north pole, when it is perpendicular. As the current directly determines the change of $\psi$, we can easily sketch flow lines on the circle in figure~\ref{fig:symbreak_locking}(a) indicating the stable fixed point, where locking will occur.  
In figure~\ref{fig:symbreak_locking}(c), we show a locked solutions of the dynamical simulation governed by the full Hamiltonian \ref{eq:sm_H_SBL_Bessel_a} without the $\alpha_0 \ll 1$ approximation which conforms with the analytical calculation.   

\section{Conclusion and Outlook}
Squeezed states of light can be created in a microwave cavity by dc-biasing a Josephson junction connected in series to it. 
When applying a dc-voltage at the parametric resonance of a single cavity, $2e V_\mathrm{dc} \approx 2 \hbar \omega_a$, the Josephson junction's nonlinearity will create pairs of photons in that cavity, while in a two-mode setup, a resonant bias with $2e V_\mathrm{dc} \approx \hbar (\omega_a + \omega_b)$ yields two-mode squeezed states. The squeezed states are reflected in steady-state distributions showing a $\mathbb{Z}^{k=2}$ symmetry in the respective phase space, with a well-defined squeezing angle and a nonzero steady-state expectation value $\braket{\ann^2}_\mathrm{st}$. 
Since dc-biasing lacks a reference phase, this simple way of creating squeezed light however suffers from a phase instability stemming from the voltage noise present in any realistic experimental setup. 
Here, we first have included a small in-series resistance $R_0$ in the theory modeling the circuit, which allows us to efficiently simulate the effect of shot noise of Cooper pairs tunneling through the junction. This captures how noise hides many (otherwise experimentally observable) properties of squeezing by effectively averaging over the neutrally stable squeezing angle, resulting in rotationally symmetric phase-space distribution. Second, we  have applied injection locking to restore the ideally expected phase-space distributions: two methods of injection locking are possible, either by adding a small ac-voltage signal to the dc-drive or by directly injecting a microwave into the cavity. Both methods create an effective sinusoidal potential for the phase-space angle $V(\psi)$, which can be trapped in one of the potential's minima and is thus protected against noise. 

For locking via an ac-bias, we find that ac-assisted Cooper pair tunneling stabilizes the current in a Shapiro-like manner. The $\mathbb{Z}^{k=2}$ symmetry of the system is protected in this case, if the applied signal is also at the parametric resonance, $\hbar\Omega \approx 2e V_\mathrm{dc}$.
In contrast to the case where the system is biased at the single-photon resonance, where an observable hallmark of locking is the re-sharpening of the noise-broadened emission spectrum of the cavity, for squeezing even the ideal spectrum is broadened.
An experimentally accessible observable of the stabilization of squeezing is the incomplete decay of the phase-phase correlation function $G^\phi$ to a finite, nonzero expectation value $|\braket{\ann^2}_\mathrm{st}|^2$. 
 Large zero-point fluctuations of the cavity can drastically change the strength and sign of the created potential. Then, the stabilization of the current by ac-assisted Cooper pair tunneling is strongly modified by the presence of the cavity, which allows an exchange of photons and additionally adds a higher-order squeezing term.
 
Injection locking through a direct microwave drive, in contrast, will not have Shapiro-like terms. Although breaking the $\mathbb{Z}^{k=2}$ symmetry, the method may be experimentally attractive, as symmetry breaking can be kept small or completely removed by an additional phase-space shift.
We conclude that the proposed locking mechanism works with both methods of injection, thus overcoming a crucial impediment for employing dc-driven Josephson-photonics devices, a quantum source lacking phase stability, for quantum information transfer and processing technologies. 

Beyond their immediate impact, our results allow more generic conclusions: The model and method applied here for the locking of various squeezed states can immediately be applied to any other phase-sensitive quantum state, which can be created in Josephson-photonics devices by accessing different resonances by the dc-bias: e. g., for Schr\"odinger cat states obtainable at a difference-resonance of a two-mode setup \cite{aissaoui2024cat}. The classical locking theory \cite{danner2021injection} has very recently been applied to such a scenario \cite{aissaoui2024cat}.
Based on the insights gained here, conclusions on the specifics of the locking mechanism, its efficiency and parameter dependence
can be drawn directly from deriving the RWA-Hamiltonian for the respective injection scheme and resonance.
Based on our classical investigations \cite{danner2021injection} and results of \cite{hoehe2023quantum} for near coherent states at the fundamental resonance it is also expected that several Josephson-photonics devices creating squeezed states can mutually synchronize their emission in a completely analogous way in which they lock onto an external ac-signal \--- exemplifying the close relation between the respective universal phenomenological Adler and Kuramoto model \cite{Adler1964,Kuramoto1975,Acebron,Pikovsky_Rosenblum_Kurths_2001}. Notably, this would dispense with the need for any external ac-generator and allow extremely compact setups.
Finally, this work may give an impulse to investigate feedback-locking of a quantum state for other setups lacking an ac-drive in completely different platforms, e.g. for optical frequencies. 

One more reason, why making dc-driven quantum microwaves sources fit for application is so important, is their perfect suitability as an on-chip component of a larger integrated device. Proposals for an on-chip coherent source \cite{Yan2021} and very recently of a microwave comb \cite{wang2024integrated} emphasize compactness and energy efficiency of such an approach. While overcoming cooling power and refrigerator size restrictions may be immediate advantages, there is also a growing interest in the resource efficiency of quantum information processing devices in general, both for practical and for fundamental reason \cite{jaschke2023quantum,Whitney_2023,smierzchalski2024efficiency,ikonen2017energy}. Requiring only minute ac-power for locking or none for synchronization, this work shows that integrated on-chip sources can not only yield coherent classical microwave radiation, but also squeezed states or other phase-sensitive quantum states.

\section{Acknowledgements}
We thank Andrew Armour, Max Hofheinz, Benjamin Huard, Ambroise Peugeot, Thiziri Aissaoui and Anil Murani for fruitful discussions and acknowledge the support of the DFG through AN336/17-1 and AN336/18-1 and the BMBF through QSolid.

%% file: text_supp.tex
\section{Hamiltonians and Cooper-Pair Currents in Rotating-Wave Approximation}
\label{sec:ap_rwa}

 In this appendix, we provide the derivation for the rotating-wave Hamiltonians and Cooper pair currents for all cases studied in the main text. 

\subsection{Single Mode: Symmetry-Preserving Locking}
\label{sec:ap_rwa_sm_sympres}

The circuit from figure \ref{fig:singlemode_wigner}, consisting of a single-mode cavity connected in series with a Josephson junction and a resistor $R_0$, where a voltage bias ${V(t)=V_\mathrm{dc} + V_\mathrm{ac}\cos(\phi_{ac})}$ with $\phi_{ac}=\Omega t + \varphi_\mathrm{ac}$ is applied, can be described by the Hamiltonian
\begin{equation}
    \label{eq:singlemode_raw}
    \hat{H} = \hbar \omega_a \adag \ann - E_J \cos\left[\omega_\mathrm{dc} t  + \varepsilon \sin(\phi_{ac}) + \alpha_0(\adag + \ann)  - \varphi_{R_0} \right].
\end{equation}
Here, we have defined $\varepsilon = \frac{2e}{\hbar \Omega}V_\mathrm{ac}$ and $\omega_\mathrm{dc} =  \frac{2e}{\hbar}V_\mathrm{dc}$. We assume that the dc-voltage is applied at the two-photon resonance $\omega_\mathrm{dc} \approx 2 \omega_a$ and that the frequency of the locking signal is applied at $\Omega \approx \omega_\mathrm{dc}$. 
We move to a rotating reference frame defined by the unitary transformation
\begin{equation}
    \hat{U}_a = \exp[i \phi_\mathrm{rot,a} \adag \ann] = \exp[i (\omega_\mathrm{rot,a} t + \phi_a) \adag \ann]
\end{equation}
and linearize in the locking strength $\varepsilon$. Then, we perform a rotating wave approximation (keeping only slowly oscillating terms), which yields 
\begin{equation}
\label{eq:singlemode_rwa_rot} 
\hat{H}_\mathrm{rwa} = \hbar (\omega_a - \omega_\mathrm{rot,a}) \adag \ann + E_J^* (\hat{h} + \hat{h}^\dag)    
\end{equation}
with 
\begin{equation}
    \label{eq:singlemode_rwa_rot_hdag}
    \eqalign{
    \hat{h}^\dag =
    & \frac{\alpha_0^2}{4} e^{i[2 \phi_\mathrm{rot,a} - (\omega_\mathrm{dc} t - \varphi_{R_0})]} : \adag[2] \frac{J_2(2\alpha_0^2 \sqrt{\adag \ann})\cdot 2}{(\alpha_0 \sqrt{\adag \ann})^2}:\cr 
    &+ \frac{\varepsilon}{4} e^{i[\phi_\mathrm{ac} - (\omega_\mathrm{dc} t - \varphi_{R_0})]} : J_0(2\alpha_0 \sqrt{\adag \ann}):\cr
    &- \frac{\varepsilon}{4}\frac{\alpha_0^4}{4!} e^{i[4\phi_\mathrm{rot,a} - \phi_\mathrm{ac} - (\omega_\mathrm{dc} t - \varphi_{R_0})]} : \adag[4] \frac{J_4(2\alpha_0^2 \sqrt{\adag \ann})\cdot 4!}{(\alpha_0 \sqrt{\adag \ann})^4}: \,\, .
    }  
\end{equation}
Here, $E_J^* = E_J \cdot e^{-\alpha_0^2/2}$ is the renormalized Josephson energy and the colons signal normal ordering of operators. 
Since the Cooper-pair current is defined as
\begin{equation}
    \label{eq:singlemode_raw_current}
    \hat{I}_J =  I_\mathrm{crit} \sin\left[\omega_\mathrm{dc} t  + \varepsilon \sin(\phi_{ac}) + \alpha_0(\adag + \ann)  - \varphi_{R_0} \right]
\end{equation}
with the critical current $I_\mathrm{crit}=\frac{2e}{\hbar} E_J$, the dc-component in rotating wave approximation is 
\begin{equation}
\hat{I}_\mathrm{CP} = \frac{2e}{\hbar} E_J^* (i \hat{h} -i \hat{h}^\dag) \, .     
\end{equation}
In approximation up to order $\alpha_0^2$, note that $J_0(2\alpha_0 \sqrt{\adag \ann}) \approx \mathbf{1} - \alpha_0^2 \adag\ann$. Further, we can neglect the fraction with the Bessel function in the first term and the complete last term in equation (\ref{eq:singlemode_rwa_rot_hdag}).

\subsection{Two Modes: Symmetry-Preserving Locking}
\label{sec:ap_rwa_tm_sympres}

We now consider two cavities connected in series with a Josephson junction and an additional resistor $R_0$. We assume that the cavities are non-degenerate $|\omega_a -\omega_b |\gg \gamma_{a/b}$. The circuit is dc-biased at the two-mode resonance $\omega_\mathrm{dc} \approx \omega_a + \omega_b$ and a locking signal with $\Omega \approx \omega_{dc}$ is applied. Starting from the lab-frame Hamiltonian   
\begin{equation}
    \fl
    \hat{H} = \hbar \omega_a \adag \ann + \hbar \omega_b \bdag \bnn - E_J \cos\left[\omega_\mathrm{dc} t + \varepsilon \sin(\phi_\mathrm{ac}) + \alpha_0(\adag + \ann) + \beta_0(\bdag + \bnn) - \varphi_{R_0} \right]
\end{equation}
we again move to a rotating frame with $\hat{U} = \hat{U}_a \otimes \hat{U}_b$ (with $\omega_\mathrm{rot,a} + \omega_\mathrm{rot,b} = \omega_\mathrm{rot, tot}$  such that $\omega_\mathrm{rot,a/b} \approx \omega_{a/b}$). Then only the sum of the (single-space) rotating frame frequencies is fixed and we choose $\delta = \omega_a - \omega_\mathrm{rot,a} =  \omega_b - \omega_\mathrm{rot,b}$. 
In rotating wave approximation, we then find 
\begin{equation}
\hat{H}_{rwa}= \hbar \delta \frac{\adag \ann + \bdag \bnn}{2} + E_J^* (\hat{h} + \hat{h}^\dag)    
\end{equation}
and 
\begin{equation}
\hat{I}_\mathrm{CP} = \frac{2e}{\hbar}E_J^* (i\hat{h} - i \hat{h}^\dag) \, , 
\end{equation}
where now  $E_J^* = E_J e^{-(\alpha_0^2 + \beta_0^2)/2}$ and     
\begin{equation}
   \label{eq:ap_twomode_rwa_hdag}
   \fl
   \eqalign{
    \hat{h}^\dag = 
    &\frac{\alpha_0 \beta_0}{2} e^{i[\phi_\mathrm{rot,a} + \phi_\mathrm{rot,b} - (\omega_\mathrm{dc}t - \varphi_{R_0})]} : \adag\bdag \frac{J_1(2\alpha_0 \sqrt{\adag \ann})}{\alpha_0 \sqrt{\adag \ann}}\frac{J_1(2\beta_0 \sqrt{\bdag \bnn})}{\beta_0 \sqrt{\bdag \bnn}}:\cr
    &+ \frac{\varepsilon}{4} e^{i[\phi_{ac} -(\omega_\mathrm{dc}t - \varphi_{R_0})]} :J_0(2\alpha_0 \sqrt{\adag \ann}) J_0(2\beta_0 \sqrt{\bdag \bnn}):\cr
    &- \frac{\varepsilon}{4} \frac{\alpha_0^2 \beta_0^2}{4} e^{i[2\phi_\mathrm{rot,a} + 2\phi_\mathrm{rot,b} - \phi_{ac} - (\omega_\mathrm{dc}t - \varphi_{R_0})]} : \adag[2] \bdag[2] \frac{J_2(2\alpha_0 \sqrt{\adag \ann})\cdot 2}{(\alpha_0 \sqrt{\adag \ann})^2}\frac{J_2(2\beta_0 \sqrt{\bdag \bnn})\cdot2}{(\beta_0 \sqrt{\bdag \bnn})^2}: \,\,. 
    }
\end{equation}

Keeping only maximally quadratic powers of zero-point fluctuations, we can neglect the fractions of Bessel functions in the first term and the complete last term of (\ref{eq:ap_twomode_rwa_hdag}), while approximating $:J_0(2\alpha_0 \sqrt{\adag \ann}) J_0(2\beta_0 \sqrt{\bdag \bnn}): \,\,\approx \mathbf{1} - \alpha_0^2 \adag \ann - \beta_0^2 \bdag \bnn$.

\subsection{Single Mode: Symmetry Breaking Locking}
\label{sec:ap_symbreak_locking}

 We consider a single mode biased at the two-photon resonance $\omega_\mathrm{dc} \approx 2 \omega_a$ of a dc-biased Josephson photonics circuit, similar to figure \ref{fig:singlemode_wigner} in the main text. Now however, the locking signal is not applied by an ac-voltage, but is directly injected as a microwave locking signal with an additional drive term $\hat{H}_\mu = \varepsilon_\mu \cos(\Omega t + \phi_\mu) \cdot (\adag + \ann)$ with effective strength $\varepsilon_\mu$ and frequency $\Omega \approx \omega_a$ into the cavity. The constant phase $\phi_\mu$ is determined by the coupling of the injected microwave signal. With an analogous rotating-frame transformation with $\omega_\mathrm{rot,a} = \Omega$ and rotating wave approximation as in \ref{sec:ap_rwa_sm_sympres}, we find
 \begin{equation}
 \label{eq:sm_H_SBL_Bessel_a}
  \hat{H}_\mathrm{rwa} = \hbar (\omega_a -\Omega) + E_J^* (\hat{h} + \hat{h}^\dag) + \frac{\varepsilon_\mu}{2}(\adag e^{i \phi_\mu} + \ann e^{-i \phi_\mu})   
 \end{equation}
 with 
 \begin{equation}
 \label{eq:sm_H_SBL_Bessel_b}
    \hat{h}^\dag = \frac{\alpha_0^2}{4} e^{i[2 \phi_\mathrm{rot,a} - (\omega_\mathrm{dc} t - \varphi_{R_0})]} : \adag[2] \frac{J_2(2\alpha_0^2 \sqrt{\adag \ann})\cdot 2}{(\alpha_0 \sqrt{\adag \ann})^2}:  \, \, . 
 \end{equation}

 We can apply a operator transformation $\ann = \alpha \mathbf{1} + \bnn$, which is effectively a constant shifts the frame of the phase-space. If the cavity is coherently driven and damped ($E_J = 0$), its steady state will be a coherent state (which is a displaced vacuum state) with steady-state amplitude $\braket{a}_{st}$. An operator transformation with $\alpha=\braket{\ann}_{st}$ effectively shifts the center of the phase-space to $\braket{\ann}_{st}$. Under this transformation, mode $\bnn$ is undriven and damped, yielding a vacuum state as a steady state. 
 
 In the limit of small Bessel function argument, the Hamiltonian can be approximated to 
 \begin{equation}
     \hat{H}_\mathrm{rwa} \approx \hbar \delta \adag \ann +\frac{E_J^*\alpha_0^2}{4} (e^{-i\psi} \ann^2 + e^{i\psi} \adag[2]) + \frac{\varepsilon_\mu}{2}(\adag e^{i \phi_\mu} + \ann e^{-i \phi_\mu}) \, , 
 \end{equation}
 where $\delta=\omega_a - \Omega$ and $\psi = 2 \phi_\mathrm{rot,a} - (\omega_\mathrm{dc}t - \varphi_{R_0})$. This bilinear system consists of a Hamiltonian with a squeezing drive and coherent drive, amended by a Lindblad operator $\hat{L}_0 = \sqrt{\gamma_a} \ann$. It can be solved analytically for constant $\psi$. If the same operator transformation is applied, the system is analogously described by a Hamiltonian 
 \begin{equation}
 \label{eq:SMl_H_RWA_SBL_trafo}
    \fl
    \eqalign{
    \hat{H}_\mathrm{rwa} \approx &\hbar \delta \bdag \bnn +\frac{E_J^*\alpha_0^2}{4} (e^{-i\psi} \bnn^2 + e^{i\psi} \bdag[2])\cr 
    &+ \left[\bdag \left(\frac{\varepsilon_\mu}{2}e^{i \phi_\mu}  -i\hbar \frac{\alpha}{2} (\gamma_a + 2i\delta) + \alpha^* \frac{E_J^* \alpha_0^2}{2} e^{i \psi} \right) + \mathrm{h.c.} \right]}
 \end{equation}
 and a Lindblad operator  $\hat{L}_b = \sqrt{\gamma_a} \bnn$. 
 The linear drive term vanishes if 
 \begin{equation}
    \varepsilon_\mu e^{i \phi_\mu}  = i \hbar \gamma_a \alpha \frac{\gamma_a + 2i\delta}{\gamma_a} -  \alpha^*E_J^* \alpha_0^2 e^{i \psi}  \, . 
 \end{equation}
  Again, the condition is fulfilled for the steady-state solution, $\alpha=\braket{\ann}_{st}$. Then, mode $\bnn$ behaves like a squeezed damped single mode and the Cooper-pair current can be described by
 \begin{equation}
     \hat{I}_\mathrm{CP} \approx \frac{2e}{\hbar} \frac{E_J^*\alpha_0^2}{4} i e^{-i\psi}\left(\bnn^2 + \alpha^2 \mathbf{1} \right) + \mathrm{h.c.} \, \, . 
 \end{equation}
 If eg. $\delta=0$, we find
 \begin{equation}
  \label{eq:ap_alpha_solution}
     \alpha =  r_\mathrm{c} e^{i (\phi_\mu - \pi/2)} + r_\mathrm{ell} e^{i (\psi - \phi_\mu)} \, .
 \end{equation}
 Varying $\psi$, the squeezing ellipse of mode $\bnn$ performs a circular motion $e^{i(\psi - \phi_\mu)}$ with radius 
 \begin{equation}
   r_\mathrm{ell} = \frac{\frac{\varepsilon_\mu}{\hbar \gamma_a}}{1 - \left(\frac{E_J^*\alpha_0^2}{\hbar \gamma_a}\right)^2} \frac{E_J^*\alpha_0^2}{\hbar \gamma_a}    
 \end{equation}
 about a new center shifted by the amount 
 \begin{equation}
   r_\mathrm{c} = \frac{\frac{\varepsilon_\mu}{\hbar \gamma_a}}{1 - \left(\frac{E_J^*\alpha_0^2}{\hbar \gamma_a}\right)^2}    
 \end{equation}
 The orientation of the ellipse (direction of the major axis) is $\phi_\mathrm{ell} = \psi/2 - \pi/4$.  

\subsection{Matrix Elements of Normal-Ordered Bessel Functions}
\label{sec:ap_laguerre_mat_elem}

 Using the series expansion of Bessel functions and the properties of the creation and annihilation operators, we find for $k, n \in \mathbf{N}_0$ the only nonzero matrix elements
 \begin{equation}
     \bra{n+k} :\adag[k] \cdot \frac{J_k(2\alpha_0 \sqrt{\adag \ann}) \cdot k! }{(\alpha_0 \sqrt{\adag \ann})^k}: \ket{n} = \sqrt{\frac{(n+k)!}{n!}} \cdot \frac{L_{n}^{(k)}(\alpha_0^2)}{L_{n}^{(k)}(0)}
 \end{equation}
 An applied dc-voltage with $\omega_\mathrm{dc} \approx 2\omega_a$ then eg. drives the two-photon transition, where in the Hamiltonian matrix elements $\bra{n+2} \hat{H} \ket{n}$ the nonlinear effect of the zero-point fluctuations is expressed by Laguerre polynomials $L_n^{(2)}(\alpha_0^2)$.

\section{Two-Time Perturbation Theory}
\label{sec:ap_tt}

In \cite{hoehe2023quantum}, two-time perturbation theory was utilized to obtain an equation of motion for the reduced dynamics of the phase $\psi$. Here, we want to extend this approach to calculate the full quantum state of the system and to obtain two-time correlation functions.

\subsection{Reduced dynamics of the phase}
As the dynamics of the cavity and the phase $\psi$ takes place on two separate time-scales given by $\gamma_a$ and $r_0\gamma_a$ we explicitly introduce the fast time $t$ and the slow time $\tau= r_0 t$. In the limit $t\to \infty$ where fast oscillator dynamics have already relaxed, the time evolution of the system is given by
\begin{equation}
    \label{eq:tteom}
    r_0\partial_\tau\sket{\rho(\tau, \psi)} = \Lambda(\psi) \sket{\rho(\tau, \psi)},
\end{equation}
with the generator of the time-evolution $\Lambda(\psi) = \mathcal{L} + r_0 \partial_\psi \Lambda^{(1)} + \mathcal{O}(r_0^2)$ where $\mathcal{L}$ is the Liouvillian of the system and $\Lambda^{(1)} = \delta_\mathrm{ac}/r_0 - 2\pi \hat{I}_\mathrm{CP}/(2e)$. By expanding $\sket{\rho(\tau, \psi)} = \sum_{l=0}^\infty r_0^l \hat{\rho}^{(l)}$ we can solve \ref{eq:tteom} order by order in $r_0$. Accordingly we define $P(\tau, \psi) = \sum_{l=0}^\infty r_0^l P^{(l)}(\tau, \psi) = \sum_{l=0}^\infty r_0^l\, \mathrm{tr}\, \hat{\rho}^{(l)}(\tau, \psi)$. In \textit{zeroth order} we find 
\begin{equation}
    0 = \mathcal{L}(\psi) \sket{\rho^{(0)}(\tau, \psi)},
\end{equation}
which determines $\hat{\rho}^{(0)}(\tau, \psi) = P^{(0)}(\tau, \psi) \hat{\rho}_\mathrm{eq}(\psi)$ up to the unknown factor $P^{(0)}(\tau, \psi)$ with $\mathrm{tr}\, \rho_\mathrm{eq}(\psi)=1$.\\
To find $P^{(0)}(\tau, \psi)$ we consider the \textit{first order} equation
\begin{equation}
    \label{eq:firstorder}
    \partial_\tau \sket{\rho^{(0)}(\tau, \psi)} = \mathcal{L}(\psi) \sket{\rho^{(1)}(\tau, \psi)} + \partial_\psi\left[\Lambda^{(1)}(\psi) \sket{\rho^{(0)}(\tau, \psi)}\right].
\end{equation}
Taking the trace, i.e. multiplying $\sbra{1}$ from the left, we find the desired equation of motion
\begin{equation}
    \partial_\tau P^{(0)}(\tau, \psi) = \partial_\psi \left[ \sbra{1}\Lambda^{(1)}(\psi)\sket{\rho_\mathrm{eq}(\psi)}P^{(0)}(\tau, \psi)\right].
\end{equation}
We note, that we did not extract the full information of (\ref{eq:firstorder}) by taking its trace. To do so, we define $\mathcal{R}$ as the pseudo-inverse of $\mathcal{L}$, i.e. $\mathcal{R} = (\mathcal{L} + \mathcal{R})^{-1} - \mathcal{P}$ with $\mathcal{P} = \sket{\rho_\mathrm{eq}}\sbra{1}$ whereby
\begin{equation}
    \mathcal{L}\mathcal{R} = \mathcal{R}\mathcal{L} = \mathcal{Q}
\end{equation}
with $\mathcal{Q} = \mathbf{1} - \mathcal{P}$ applies. Then multiplying (\ref{eq:firstorder}) with $\mathcal{R}$ from left yields
\begin{equation}
    0 = \mathcal{R}\mathcal{L}\sket{\rho^{(1)}} + \mathcal{R}\partial_\psi\left(\Lambda^{(1)}\sket{\rho^{(0)}}\right)
\end{equation}
and equivalently
\begin{equation}
    \label{eq:uselater}
    \mathcal{Q}\sket{\rho^{(1)}} = - \mathcal{R}\partial_\psi\left(\Lambda^{(1)}\sket{\rho^{(0)}}\right).
\end{equation}
which we will need for the full reconstruction of the quantum state.\\

As shown in \cite{hoehe2023quantum} the \textit{second order} of (\ref{eq:tteom}) yields a Fokker-Planck equation for $P(\psi) \approx P^{(0)}(\tau, \psi) + r_0 P^{(1)}(\tau, \psi)$ 
\begin{equation}
    \label{eq:fp}
    \partial_\tau P(\tau, \psi) = -\partial_\psi [j(\psi)P(\tau, \psi)] + \partial_\psi^2[D(\psi)P(\tau, \psi)]
\end{equation}
with
\begin{equation}
    \label{eq:j0}
    j^{(0)}(\psi) = -\sbra{1}\Lambda^{(1)}\sket{\rho_\mathrm{eq}}
\end{equation}
\begin{equation}
    \label{eq:j1}
    j^{(1)}(\psi) = -\sbra{1}\partial_\psi\left(\Lambda^{(1)}\mathcal{R}\right) \Lambda^{(1)}\sket{\rho_\mathrm{eq}}
\end{equation}
\begin{equation}
    \label{eq:D}
    D(\psi) = - \sbra{1}\Lambda^{(1)} \mathcal{R} \Lambda^{(1)}\sket{\rho_\mathrm{eq}},
\end{equation}
containing shot noise of the CP current \cite{Armour2017}.

\subsection{Reconstruction of the state}
Now, we want to go one step further and find the full density matrix $\hat{\rho}(\tau) = \int_{0}^{2\pi} \mathrm{d}\psi \, \hat{\rho}(\tau, \psi)$. In zeroth order we have
\begin{equation}
    \sket{\rho^{(0)}(\tau, \psi)} = P^{(0)}(\tau, \psi)\sket{\rho_\mathrm{eq}(\psi)}
\end{equation}
in the first order
\begin{equation}
    \fl
    \eqalign{
    \sket{\rho^{(1)}(\tau, \psi)} &= \left(\mathcal{P}(\psi) + \mathcal{Q}(\psi)\right)\sket{\rho^{(1)}(\tau, \psi)}\cr 
    &= \sket{\rho_\mathrm{eq}(\psi)}\sbraket{1}{\rho^{(1)}(\tau, \psi)} + \mathcal{Q}(\psi)\sket{\rho^{(1)}(\tau, \psi)}\cr
    &= P^{(1)}(\tau, \psi)\sket{\rho_\mathrm{eq}(\psi)} - \mathcal{R}(\psi) \partial_\psi\left(\Lambda^{(1)}(\psi)\sket{\rho^{(0)}(\tau, \psi)}\right)
    }
\end{equation}
Summing up both orders we obtain
\begin{equation}
    \fl
    \eqalign{
    \sket{\rho(\tau, \psi)} \approx \sket{\rho^{(0)}(\tau, \psi)} + r_0 \sket{\rho^{(1)}(\tau, \psi)}\cr
    = \left((P^{(0)}(\tau, \psi) + r_0 P^{(1)}(\tau, \psi)\right)\sket{\rho_\mathrm{eq}(\psi)} - r_0 \mathcal{R}(\psi)\partial_\psi \left(\Lambda^{(1)}(\psi)  P^{(0)}(\tau, \psi)\sket{\rho_\mathrm{eq}(\psi)}\right)\cr
    = P(\tau, \psi)\sket{\rho_\mathrm{eq}(\psi)} - r_0\mathcal{R}(\psi) \partial_\psi \left(\Lambda^{(1)}(\psi)  P(\tau, \psi)\sket{\rho_\mathrm{eq}(\psi)}\right).}
\end{equation}

\subsection{Fast timescale dynamics and two-time correlation functions}
In \textit{zeroth order} order of $r_0$ the fast time-scale dynamics is given by
\begin{equation}
    \label{eq:fast}
    \partial_t \sket{\rho_\mathrm{eq}(t, \psi)} = \Lambda^{(0)}(\psi)\sket{\rho_\mathrm{eq}(t, \psi)},
\end{equation}
which yields the full density matrix of the system
\begin{equation}
    \hat{\rho}(t, \psi) = P(\tau, \psi)\hat{\rho}_\mathrm{eq}(t, \psi), 
\end{equation}
where $P(\tau, \psi)$ is evolved via the Fokker-Planck equation (\ref{eq:fp}). 

To calculate two-time correlation functions we make use of the quantum regression theorem
\begin{equation}
    \braket{\hat{B}(t_2)\hat{A}(t_1)} = \mathrm{tr}\left\{\hat{B} \left[\hat{A}\hat{\rho}(t_1)\right]_{t_2}\right\} = \int_{0}^{2\pi} \mathrm{d}\psi \, \mathrm{tr}\left\{\hat{B}\left[\hat{A}\hat{\rho}(t_1, \psi)\right]_{t_2}\right\}
\end{equation}
To evolve the matrix $\hat{A}\rho(t_1, \psi)$ up to $t_2$, we first normalize it according to
\begin{equation}
    \hat{\rho}'(t_1, \psi) = \frac{\hat{A} \hat{\rho}(t_1, \psi)}{\mathrm{tr}\left[\hat{A} \hat{\rho}(t_1, \psi)\right]} \qquad \mathrm{and} \qquad P'(t_1, \psi) = \mathrm{tr}\left[\hat{A} \hat{\rho}(t_1, \psi)\right], 
\end{equation}
note that $P'(t_1, \psi)$ not real in general. Then we evolve $P'$ with the Fokker-Planck equation and $\rho'$ according to (\ref{eq:fast}) and get
\begin{equation}
    \braket{\hat{B}(t_2)\hat{A}(t_1)} = \int_{0}^{2\pi} \mathrm{d}\psi \, \mathrm{tr}\left[\hat{B} P'(t_2, \psi)\hat{\rho}'(t_2, \psi)\right].
\end{equation}
The time-evolution $P'$ is a diffusion process in the complex plane.

%% file: StabilizingSqueezedMWLight.bbl
\begin{thebibliography}{10}

\bibitem{xiao1987precision}
Min Xiao, Ling-An Wu, and H~Jeffrey Kimble.
\newblock Precision measurement beyond the shot-noise limit.
\newblock {\em Physical Review Letters}, 59(3):278, 1987.

\bibitem{grangier1987squeezed}
Philippe Grangier, RE~Slusher, B~Yurke, and A~LaPorta.
\newblock Squeezed-light--enhanced polarization interferometer.
\newblock {\em Physical Review Letters}, 59(19):2153, 1987.

\bibitem{barsotti2018squeezed}
Lisa Barsotti, Jan Harms, and Roman Schnabel.
\newblock Squeezed vacuum states of light for gravitational wave detectors.
\newblock {\em Reports on Progress in Physics}, 82(1):016905, 2018.

\bibitem{casariego2022}
Mateo Casariego, Emmanuel~Zambrini Cruzeiro, Stefano Gherardini, Tasio
  Gonzalez-Raya, Rui André, Gonçalo Frazão, Giacomo Catto, Mikko Möttönen,
  Debopam Datta, Klaara Viisanen, Joonas Govenius, Mika Prunnila, Kimmo
  Tuominen, Maximilian Reichert, Michael Renger, Kirill~G Fedorov, Frank Deppe,
  Harriet van~der Vliet, A~J Matthews, Yolanda Fernández, R~Assouly,
  R~Dassonneville, B~Huard, Mikel Sanz, and Yasser Omar.
\newblock Propagating quantum microwaves: towards applications in communication
  and sensing.
\newblock {\em Quantum Science and Technology}, 8(2):023001, 2023.

\bibitem{di2015quantum}
R~Di~Candia, KG~Fedorov, L~Zhong, S~Felicetti, EP~Menzel, M~Sanz, F~Deppe,
  A~Marx, R~Gross, and Enrique Solano.
\newblock Quantum teleportation of propagating quantum microwaves.
\newblock {\em EPJ Quantum Technology}, 2(1):1--17, 2015.

\bibitem{fedorov2021experimental}
Kirill~G Fedorov, Michael Renger, Stefan Pogorzalek, Roberto Di~Candia, Qiming
  Chen, Yuki Nojiri, Kunihiro Inomata, Yasunobu Nakamura, Matti Partanen, Achim
  Marx, et~al.
\newblock Experimental quantum teleportation of propagating microwaves.
\newblock {\em Science Advances}, 7(52):eabk0891, 2021.

\bibitem{gonzalez2022open}
Tasio Gonzalez-Raya, Mateo Casariego, Florian Fesquet, Michael Renger, Vahid
  Salari, Mikko M{\"o}tt{\"o}nen, Yasser Omar, Frank Deppe, Kirill~G Fedorov,
  and Mikel Sanz.
\newblock Open-air microwave entanglement distribution for quantum
  teleportation.
\newblock {\em Physical Review Applied}, 18(4):044002, 2022.

\bibitem{fesquet2024demonstration}
Florian Fesquet, Fabian Kronowetter, Michael Renger, Wun~Kwan Yam, Simon
  Gandorfer, Kunihiro Inomata, Yasunobu Nakamura, Achim Marx, Rudolf Gross, and
  Kirill~G Fedorov.
\newblock Demonstration of microwave single-shot quantum key distribution.
\newblock {\em Nature Communications}, 15(1):7544, 2024.

\bibitem{fesquet2023perspectives}
Florian Fesquet, Fabian Kronowetter, Michael Renger, Qiming Chen, Kedar
  Honasoge, Oscar Gargiulo, Yuki Nojiri, Achim Marx, Frank Deppe, Rudolf Gross,
  et~al.
\newblock Perspectives of microwave quantum key distribution in the open air.
\newblock {\em Physical Review A}, 108(3):032607, 2023.

\bibitem{kronowetter2023quantum}
F~Kronowetter, F~Fesquet, M~Renger, K~Honasoge, Y~Nojiri, K~Inomata,
  Y~Nakamura, A~Marx, R~Gross, and KG~Fedorov.
\newblock Quantum microwave parametric interferometer.
\newblock {\em Physical Review Applied}, 20(2):024049, 2023.

\bibitem{lloyd2008enhanced}
Seth Lloyd.
\newblock Enhanced sensitivity of photodetection via quantum illumination.
\newblock {\em Science}, 321(5895):1463--1465, 2008.

\bibitem{tan2008quantum}
Si-Hui Tan, Baris~I Erkmen, Vittorio Giovannetti, Saikat Guha, Seth Lloyd,
  Lorenzo Maccone, Stefano Pirandola, and Jeffrey~H Shapiro.
\newblock Quantum illumination with gaussian states.
\newblock {\em Physical Review Letters}, 101(25):253601, 2008.

\bibitem{Karsa_2024}
Athena Karsa, Alasdair Fletcher, Gaetana Spedalieri, and Stefano Pirandola.
\newblock Quantum illumination and quantum radar: a brief overview.
\newblock {\em Reports on Progress in Physics}, 87(9):094001, 2024.

\bibitem{assouly2023quantum}
R{\'e}ouven Assouly, R{\'e}my Dassonneville, Th{\'e}au Peronnin, Audrey
  Bienfait, and Benjamin Huard.
\newblock Quantum advantage in microwave quantum radar.
\newblock {\em Nature Physics}, 19(10):1418--1422, 2023.

\bibitem{chang2019}
C.~W.~Sandbo Chang, A.~M. Vadiraj, J.~Bourassa, B.~Balaji, and C.~M. Wilson.
\newblock Quantum-enhanced noise radar.
\newblock {\em Applied Physics Letters}, 114(11):112601, 2019.

\bibitem{messaoudi2020practical}
Nizar Messaoudi, CW~Sandbo Chang, AM~Vadiraj, CM~Wilson, J~Bourassa, and
  B~Balaji.
\newblock Practical advantage in microwave quantum illumination.
\newblock In {\em 2020 IEEE Radar Conference (RadarConf20)}, pages 1--5. IEEE,
  2020.

\bibitem{barzanjeh2020microwave}
Shabir Barzanjeh, Stefano Pirandola, David Vitali, and Johannes~M Fink.
\newblock Microwave quantum illumination using a digital receiver.
\newblock {\em Science advances}, 6(19):eabb0451, 2020.

\bibitem{Peichl20024}
Florian Bischeltsrieder, Michael Würth, Johannes Russer, Markus Peichl, and
  Wolfgang Utschick.
\newblock Engineering constraints and application regimes of quantum radar.
\newblock {\em IEEE Transactions on Radar Systems}, 2:197--214, 2024.

\bibitem{Didier2018}
Nicolas Didier, J\'er\'emie Guillaud, Shyam Shankar, and Mazyar Mirrahimi.
\newblock Remote entanglement stabilization and concentration by quantum
  reservoir engineering.
\newblock {\em Physical Review A}, 98:012329, 2018.

\bibitem{Eichler2011}
C.~Eichler, D.~Bozyigit, C.~Lang, M.~Baur, L.~Steffen, J.~M. Fink, S.~Filipp,
  and A.~Wallraff.
\newblock Observation of two-mode squeezing in the microwave frequency domain.
\newblock {\em Physical Review Letters}, 107:113601, 2011.

\bibitem{Bergeal2012}
N.~Bergeal, F.~Schackert, L.~Frunzio, and M.~H. Devoret.
\newblock Two-mode correlation of microwave quantum noise generated by
  parametric down-conversion.
\newblock {\em Physical Review Letters}, 108:123902, 2012.

\bibitem{flurin2012generating}
Emmanuel Flurin, Nicolas Roch, Fran{\c{c}}ois Mallet, Michel~H Devoret, and
  Benjamin Huard.
\newblock Generating entangled microwave radiation over two transmission lines.
\newblock {\em Physical Review Letters}, 109(18):183901, 2012.

\bibitem{zhong2013squeezing}
Ling Zhong, Edwin~P Menzel, Roberto Di~Candia, Peter Eder, Moritz Ihmig,
  Alexander Baust, Max Haeberlein, Erik Hoffmann, Koichiro Inomata, Takanori
  Yamamoto, et~al.
\newblock Squeezing with a flux-driven josephson parametric amplifier.
\newblock {\em New Journal of Physics}, 15(12):125013, 2013.

\bibitem{svensson2018period}
Ida-Maria Svensson, Andreas Bengtsson, Jonas Bylander, Vitaly Shumeiko, and Per
  Delsing.
\newblock Period multiplication in a parametrically driven superconducting
  resonator.
\newblock {\em Applied Physics Letters}, 113(2), 2018.

\bibitem{gu2017}
Xiu Gu, Anton~Frisk Kockum, Adam Miranowicz, Yu-xi Liu, and Franco Nori.
\newblock Microwave photonics with superconducting quantum circuits.
\newblock {\em Physics Reports}, 718:1--102, 2017.

\bibitem{wustmann2019parametric}
Waltraut Wustmann and Vitaly Shumeiko.
\newblock Parametric effects in circuit quantum electrodynamics.
\newblock {\em Low Temperature Physics}, 45(8):848--869, 2019.

\bibitem{Hofheinz2011}
Max Hofheinz, Fabien Portier, Quentin Baudouin, Philippe Joyez, Denis Vion,
  Patrice Bertet, Patrice Roche, and Daniel Est{\`e}ve.
\newblock Bright side of the coulomb blockade.
\newblock {\em Physical Review Letters}, 106(21):217005, 2011.

\bibitem{Chen2014}
Fei Chen, Juliang Li, A.~D. Armour, E.~Brahimi, Joel Stettenheim, A.~J. Sirois,
  R.~W. Simmonds, M.~P. Blencowe, and A.~J. Rimberg.
\newblock Realization of a single-cooper-pair josephson laser.
\newblock {\em Physical Review B}, 90:020506(R), 2014.

\bibitem{Cassidy2017}
M.~C. Cassidy, A.~Bruno, S.~Rubbert, M.~Irfan, J.~Kammhuber, R.~N. Schouten,
  A.~R. Akhmerov, and L.~P. Kouwenhoven.
\newblock Demonstration of an ac josephson junction laser.
\newblock {\em Science}, 355(6328):939--942, 2017.

\bibitem{Gramich2013}
Vera Gramich, Bj\"orn Kubala, Selina Rohrer, and Joachim Ankerhold.
\newblock From coulomb-blockade to nonlinear quantum dynamics in a
  superconducting circuit with a resonator.
\newblock {\em Physical Review Letters}, 111:247002, 2013.

\bibitem{Armour2013}
A.~D. Armour, M.~P. Blencowe, E.~Brahimi, and A.~J. Rimberg.
\newblock Universal quantum fluctuations of a cavity mode driven by a josephson
  junction.
\newblock {\em Physical Review Letters}, 111:247001, 2013.

\bibitem{Westig2017}
M.~Westig, B.~Kubala, O.~Parlavecchio, Y.~Mukharsky, C.~Altimiras, P.~Joyez,
  D.~Vion, P.~Roche, D.~Esteve, M.~Hofheinz, M.~Trif, P.~Simon, J.~Ankerhold,
  and F.~Portier.
\newblock Emission of nonclassical radiation by inelastic cooper pair
  tunneling.
\newblock {\em Physical Review Letters}, 119:137001, 2017.

\bibitem{Peugeot2021}
A~Peugeot, G~M{\'e}nard, Simon Dambach, M~Westig, Bj{\"o}rn Kubala,
  Y~Mukharsky, C~Altimiras, Philippe Joyez, D~Vion, P~Roche, et~al.
\newblock Generating two continuous entangled microwave beams using a dc-biased
  josephson junction.
\newblock {\em Physical Review X}, 11(3):031008, 2021.

\bibitem{Padurariu2012}
Ciprian Padurariu, Fabian Hassler, and Yuli~V. Nazarov.
\newblock Statistics of radiation at josephson parametric resonance.
\newblock {\em Physical Review B}, 86:054514, 2012.

\bibitem{Juha2013}
Juha Lepp\"akangas, G\"oran Johansson, Michael Marthaler, and Mikael
  Fogelstr\"om.
\newblock Nonclassical photon pair production in a voltage-biased josephson
  junction.
\newblock {\em Physical Review Letters}, 110:267004, 2013.

\bibitem{Armour2015}
A.~D. Armour, B.~Kubala, and J.~Ankerhold.
\newblock Josephson photonics with a two-mode superconducting circuit.
\newblock {\em Physical Review B}, 91:184508, 2015.

\bibitem{Trif2015}
Mircea Trif and Pascal Simon.
\newblock Photon cross-correlations emitted by a josephson junction in two
  microwave cavities.
\newblock {\em Physical Review B}, 92:014503, 2015.

\bibitem{Rolland2019}
C.~Rolland, A.~Peugeot, S.~Dambach, M.~Westig, B.~Kubala, Y.~Mukharsky,
  C.~Altimiras, H.~le~Sueur, P.~Joyez, D.~Vion, P.~Roche, D.~Esteve,
  J.~Ankerhold, and F.~Portier.
\newblock Antibunched photons emitted by a dc-biased josephson junction.
\newblock {\em Physical Review Letters}, 122:186804, 2019.

\bibitem{Grimm2019}
A.~Grimm, F.~Blanchet, R.~Albert, J.~Lepp\"akangas, S.~Jebari, D.~Hazra,
  F.~Gustavo, J.-L. Thomassin, E.~Dupont-Ferrier, F.~Portier, and M.~Hofheinz.
\newblock Bright on-demand source of antibunched microwave photons based on
  inelastic cooper pair tunneling.
\newblock {\em Physical Review X}, 9:021016, 2019.

\bibitem{Menard2022}
G.~C. M\'enard, A.~Peugeot, C.~Padurariu, C.~Rolland, B.~Kubala, Y.~Mukharsky,
  Z.~Iftikhar, C.~Altimiras, P.~Roche, H.~le~Sueur, P.~Joyez, D.~Vion,
  D.~Esteve, J.~Ankerhold, and F.~Portier.
\newblock Emission of photon multiplets by a dc-biased superconducting circuit.
\newblock {\em Physical Review X}, 12:021006, 2022.

\bibitem{Wang2017}
Hui Wang, M.~P. Blencowe, A.~D. Armour, and A.~J. Rimberg.
\newblock Quantum dynamics of a josephson junction driven cavity mode system in
  the presence of voltage bias noise.
\newblock {\em Physical Review B}, 96:104503, 2017.

\bibitem{hoehe2023quantum}
Florian Höhe, Ciprian Padurariu, Brecht I.~C Donvil, Lukas Danner, Joachim
  Ankerhold, and Björn Kubala.
\newblock Quantum synchronization in presence of shot noise, 2023.

\bibitem{Pikovsky_Rosenblum_Kurths_2001}
Arkady Pikovsky, Michael Rosenblum, and Jürgen Kurths.
\newblock {\em Synchronization: A Universal Concept in Nonlinear Sciences}.
\newblock Cambridge Nonlinear Science Series. Cambridge University Press, 2001.

\bibitem{Bengtson2018}
Andreas Bengtsson, Philip Krantz, Micha\"el Simoen, Ida-Maria Svensson, Ben
  Schneider, Vitaly Shumeiko, Per Delsing, and Jonas Bylander.
\newblock Nondegenerate parametric oscillations in a tunable superconducting
  resonator.
\newblock {\em Physical Review B}, 97:144502, 2018.

\bibitem{markovic2019injection}
Danijela Markovi{\'c}, Jean-Damien Pillet, Emmanuel Flurin, Nicolas Roch, and
  Benjamin Huard.
\newblock Injection locking and parametric locking in a superconducting
  circuit.
\newblock {\em Physical Review Applied}, 12(2):024034, 2019.

\bibitem{Xu2013}
Canran Xu and Maxim~G Vavilov.
\newblock Full counting statistics of photons emitted by a double quantum dot.
\newblock {\em Physical Review B}, 88(19):195307, 2013.

\bibitem{Shapiro1963}
Sidney Shapiro.
\newblock Josephson currents in superconducting tunneling: The effect of
  microwaves and other observations.
\newblock {\em Physical Review Letters}, 11:80--82, 1963.

\bibitem{Kramers1940}
Hendrik~Anthony Kramers.
\newblock Brownian motion in a field of force and the diffusion model of
  chemical reactions.
\newblock {\em Physica}, 7(4):284--304, 1940.

\bibitem{risken1996fokker}
Hannes Risken.
\newblock {\em The Fokker-Planck equation}.
\newblock Springer, 1996.

\bibitem{DoMu2018}
B.~Donvil, P.~Muratore-Ginanneschi, J.~P. Pekola, and K.~Schwieger.
\newblock {Model for calorimetric measurements in an open quantum system}.
\newblock {\em Physical Review A}, 97:052107, 2018.

\bibitem{pavliotis2008multiscale}
G.A. Pavliotis and A.~Stuart.
\newblock {\em Multiscale Methods: Averaging and Homogenization}.
\newblock Texts in Applied Mathematics. Springer New York, 2008.

\bibitem{Adler1964}
R.~{Adler}.
\newblock A study of locking phenomena in oscillators.
\newblock {\em Proceedings of the IRE}, 34(6):351--357, 1946.

\bibitem{dambach2015time}
Simon Dambach, Bj{\"o}rn Kubala, Vera Gramich, and Joachim Ankerhold.
\newblock Time-resolved statistics of nonclassical light in josephson
  photonics.
\newblock {\em Physical Review B}, 92(5):054508, 2015.

\bibitem{fedorov2018finite}
Kirill~G Fedorov, S~Pogorzalek, U~Las~Heras, M~Sanz, P~Yard, P~Eder, M~Fischer,
  J~Goetz, E~Xie, K~Inomata, et~al.
\newblock Finite-time quantum entanglement in propagating squeezed microwaves.
\newblock {\em Scientific reports}, 8(1):6416, 2018.

\bibitem{Kubala2015}
B.~Kubala, V.~Gramich, and J.~Ankerhold.
\newblock {Non-classical light from superconducting resonators coupled to
  voltage-biased Josephson junctions}.
\newblock {\em Physica Scripta}, T165:014029, 2015.

\bibitem{Kubala2020}
Björn Kubala, Joachim Ankerhold, and Andrew~D Armour.
\newblock Electronic and photonic counting statistics as probes of
  non-equilibrium quantum dynamics.
\newblock {\em New J. Phys}, 22(2):023010, 2020.

\bibitem{lang2021}
Ben Lang and Andrew~D Armour.
\newblock Multi-photon resonances in josephson junction-cavity circuits.
\newblock {\em New Journal of Physics}, 23(3):033021, 2021.

\bibitem{aissaoui2024cat}
Thiziri Aissaoui, Anil Murani, Rapha{\"e}l Lescanne, and Alain Sarlette.
\newblock A cat qubit stabilization scheme using a voltage biased josephson
  junction.
\newblock {\em arXiv preprint arXiv:2411.08132}, 2024.

\bibitem{danner2021injection}
Lukas Danner, Ciprian Padurariu, Joachim Ankerhold, and Bj\"orn Kubala.
\newblock Injection locking and synchronization in josephson photonics devices.
\newblock {\em Physical Review B}, 104:054517, 2021.

\bibitem{Kuramoto1975}
Y.~Kuramoto.
\newblock International symposium on mathematical problems in theoretical
  physics.
\newblock {\em Lecture Notes in Physics}, 30:420, 1975.

\bibitem{Acebron}
Juan~A. Acebr\'on, L.~L. Bonilla, Conrad~J. P\'erez~Vicente, F\'elix Ritort,
  and Renato Spigler.
\newblock The kuramoto model: A simple paradigm for synchronization phenomena.
\newblock {\em Reviews of Modern Physics}, 77:137--185, 2005.

\bibitem{Yan2021}
Chengyu Yan, Juha Hassel, Visa Vesterinen, Jinli Zhang, Joni Ikonen, Leif
  Gr{\"o}nberg, Jan Goetz, and Mikko M{\"o}tt{\"o}nen.
\newblock A low-noise on-chip coherent microwave source.
\newblock {\em Nature Electronics}, 4(12):885--892, 2021.

\bibitem{wang2024integrated}
Chen-Guang Wang, Wuyue Xu, Chong Li, Lili Shi, Junliang Jiang, Tingting Guo,
  Wen-Cheng Yue, Tianyu Li, Ping Zhang, Yang-Yang Lyu, et~al.
\newblock Integrated and dc-powered superconducting microcomb.
\newblock {\em Nature Communications}, 15(1):4009, 2024.

\bibitem{jaschke2023quantum}
Daniel Jaschke and Simone Montangero.
\newblock Is quantum computing green? an estimate for an energy-efficiency
  quantum advantage.
\newblock {\em Quantum Science and Technology}, 8(2):025001, 2023.

\bibitem{Whitney_2023}
Marco Fellous-Asiani, Jing~Hao Chai, Yvain Thonnart, Hui~Khoon Ng, Robert~S.
  Whitney, and Alexia Auff\`eves.
\newblock Optimizing resource efficiencies for scalable full-stack quantum
  computers.
\newblock {\em PRX Quantum}, 4:040319, 2023.

\bibitem{smierzchalski2024efficiency}
Tomasz {\'S}mierzchalski, Zakaria Mzaouali, Sebastian Deffner, and
  Bart{\l}omiej Gardas.
\newblock Efficiency optimization in quantum computing: balancing
  thermodynamics and computational performance.
\newblock {\em Scientific Reports}, 14(1):4555, 2024.

\bibitem{ikonen2017energy}
Joni Ikonen, Juha Salmilehto, and Mikko M{\"o}tt{\"o}nen.
\newblock Energy-efficient quantum computing.
\newblock {\em npj Quantum Information}, 3(1):17, 2017.

\bibitem{Armour2017}
Andrew~D. Armour, Bj\"orn Kubala, and Joachim Ankerhold.
\newblock Noise switching at a dynamical critical point in a cavity-conductor
  hybrid.
\newblock {\em Physical Review B}, 96:214509, 2017.

\end{thebibliography}
